\documentclass[aps,a4paper, preprint, superscriptaddress,preprintnumbers,floatfix,nofootinbib,amsmath,amssymb]{revtex4}
\usepackage{amstext,amssymb}
\usepackage{amsmath}
\usepackage{multirow}
\usepackage{graphicx}
\usepackage{dcolumn}
\usepackage[hyperfootnotes=false]{hyperref}
\usepackage{xspace}
\usepackage{braket}
\usepackage{color}
\usepackage{units}
\usepackage{slashed} 
\usepackage{graphicx}
\usepackage{bm}
\newcommand {\ignore}[1]{}

\def\3331{$\mathrm{SU(3)_c \times SU(3)_L \times SU(3)_R \times U(1)_{X}}$}

\newcommand{\be}{\begin{equation}}
\newcommand{\ee}{\end{equation}}
\newcommand{\bea}{\begin{eqnarray}}
\newcommand{\eea}{\end{eqnarray}}
\newcommand{\nn}{\nonumber}
\begin{document}
\title{Singlet scalar Dark matter in $U(1)_{B-L}$ models without right-handed neutrinos}

\author{Shivaramakrishna Singirala}
\email{krishnas542@gmail.com}
\affiliation{School of Physics,  University of Hyderabad, Hyderabad - 500046,  India}
\author{Rukmani Mohanta}
\email{rmsp@uohyd.ernet.in}
\affiliation{School of Physics,  University of Hyderabad, Hyderabad - 500046,  India}

\author{Sudhanwa Patra}
\email{sudhanwa@iitbhilai.ac.in}
\affiliation{Indian Institute of Technology Bhilai, GEC Campus, Sejbahar, Raipur-492015, Chhattisgarh, India}

\begin{abstract}
We investigate the phenomenology of  singlet scalar dark matter in a simple  $B-L$ gauge 
extension of the Standard Model where the dark matter particle is charged under the $U(1)_{B-L}$ symmetry. 
The non-trivial gauge anomalies are cancelled with the introduction of three exotic fermions with $B-L$ 
charges as $-4,-4,5$, instead of right-handed neutrinos $\nu_{Ri}~(i=1,2,3)$  with $B-L=-1$ in conventional 
$U(1)_{B-L}$ model. Without the need of any ad-hoc discrete symmetry, the $B-L$ charge plays a crucial 
role in stabilizing the dark matter.  We make a  comprehensive study of dark matter phenomenology in the scalar and gauge portals separately. In the gauge-mediated regime, we invoke the LEP-II constraints and dilepton limits of ATLAS on the gauge parameters. A massless physical Goldstone plays a vital role in the scalar-portal dark matter observables, becomes a unique feature of the model.  We show the mechanism of generating the light neutrino mass at one-loop level where the dark matter singlet runs in the loop.
We shed light on the semi-annihilation and finally, we comment on indirect signals in this framework.
\end{abstract}
\maketitle

\section{Introduction}
\label{sec:intro}
Although there are indirect astrophysical evidences for the existence of dark matter contributing with a relic density $\Omega \text{h}^{2} \simeq 0.12$,  about 25\% to the energy budget of the Universe \cite{Ade:2015xua}, 
still we know only very little about the nature of the dark matter. In particular the  unknowns are: what kind of particle it is, i.e., scalar, fermion or vector etc., and to which beyond the Standard Model framework it belongs to (see the recent review article~\cite{Arcadi:2017kky} for details). In this respect, models in which the difference between 
baryon and lepton number ($B-L$) is gauged, are economic extensions of the Standard Model~\cite{Jenkins:1987ue,Buchmuller:1991ce,Basso:2008iv,Emam:2007dy} (see also few earlier works in this direction~\cite{Khalil:2006yi,Iso:2009ss,Kanemura:2014rpa,Lindner:2011it,Okada:2016gsh,Okada:2016tci,Bhattacharya:2016qsg,
Biswas:2016ewm,Wang:2015saa}). One of the interesting aspects of this kind of models is that in the standard form, the presence of right-handed neutrinos and thus, the type-I seesaw mechanism for neutrino mass generation appears naturally. 
In addition, attempts have also been made within this economic extensions of SM where dark matter can be incorporated as well \cite{Klasen:2016qux,Rodejohann:2015lca,Lindner:2013awa,Okada:2010wd,Okada:2012sg,
Basak:2013cga,Sanchez-Vega:2014rka,Duerr:2015wfa,Guo:2015lxa,Dasgupta:2016odo,Modak:2016ung}. 

It is widely believed that weakly interacting massive particles (WIMPs) fulfil  the necessary criteria of dark matter, not too far from the electroweak scale, which provides 
the opportunity to test them at the current or near future direct or indirect dark matter detection experiments. One of the fundamental questions is how to address the stability 
of the dark matter. Within the gauged $B-L$ extensions of the Standard Model, the stability of the dark matter can be taken care of by imposing an extra discrete symmetry on top 
of the gauge symmetry~\cite{Okada:2010wd,Okada:2012sg,Basak:2013cga,Guo:2015lxa}. In these class of models one of the right-handed neutrinos  introduced for gauge anomaly 
cancellation is odd under the additional discrete symmetry and acts as a dark matter candidate. Attempts are also made to ensure the stability of the dark matter by choosing the appropriate $B-L$ charge of dark matter 
\cite{Lindner:2013awa,Sanchez-Vega:2014rka,Duerr:2015wfa,Rodejohann:2015lca,Dasgupta:2016odo}. There are other variants of gauged $B-L$ extension of SM, 
where the additional fermions carry exotic integer value of $B-L$ charge. 
The discussion of scalar dark matter and neutrino phenomenology have been explored in the recent works~\cite{Ma:2014qra,Ma:2015raa,Ma:2015mjd}, 
while a beautiful connection between dark matter abundance and matter-antimatter asymmetry has been investigated in 
Ref.~\cite{Dasgupta:2016odo} within WIMPy Leptogenesis. 

In this work, we attempt to study the phenomenology of a scalar dark matter within the context of gauged $B-L$ model without the introduction 
of any right-handed neutrinos, which are generally present in the conventional $B-L$ theory. The induced gauge anomalies are cancelled by assigning appropriate 
$B-L$ charges to the additional fermions. The key point to note here is that the stability of the scalar singlet dark matter is ensured by the peculiar choice of $B-L$ charges and not by introducing any ad-hoc discrete symmetry. The proposed model provides another variant of  the class of gauged $B-L$ models. Similar work on singlet scalar DM phenomenology  has been recently explored in \citep{Rodejohann:2015lca} where three right-handed neutrinos are added to make the model anomaly free and the model structure itself takes care of the stability of scalar DM. Dirac DM has also recently been investigated in a $B-L$ model \citep{Patra:2016ofq}, where four exotic fermions are added to overcome the gauge anomalies. The current model describes a new variant of $B-L$ models with a different scalar content and exotic charges assigned to the newly added fermions. Moreover, the $B-L$ charge assigned to the scalar DM and its corresponding annihilation channels, the arising parameter scan are different from the conventional $B-L$ models.

The plan of the paper is as follows. We discuss in Sec-II, a simple model with  $B-L$ gauge extension of SM without right-handed neutrinos along with the allowed solutions for 
gauge anomaly cancellation, vacuum stability criteria, perturbative unitarity constraints and the mass spectrum of the scalar sector. In Sec-III, we explore the $Z^{\prime}$-mediated dark matter phenomenology of scalar singlet in view of  relic density, direct detection and collider bounds perspective. In Sec-IV, we repeat the same for the scalar-portal, then we briefly discuss the  generation of light neutrino mass and the effect of semi-annihilation to relic density in Sec-V and VI respectively. In Sec-VII, we comment on indirect signals  followed by conclusion in Sec-VIII.

\section{The model framework}
\label{sec:model}
It is believed that the $B-L$ gauge extension of Standard Model (SM) is the simplest model one can think of from the point of view of 
a self-consistent gauge theory where the difference between baryon and lepton number is promoted to local gauge symmetry. The gauge group 
of this simplest $B-L$ model is $SU(2)_L \times U(1)_Y \times U(1)_{B-L}$,  omitting the $SU(3)_C$ structure for simplicity. 
Originally, these models are motivated to cancel the triangle gauge anomalies
\begin{equation}
\mathcal{A}_1\left[U(1)^3_{B-L}\right] ,~ 
\mathcal{A}_2\left[\mbox{(gravity)}^2 \times U(1)_{B-L}\right], 
\end{equation}
with the inclusion of right-handed neutrinos $\nu_{Ri} ~(i=1,2,3)$ having the $B-L$ charges  $-1$ (the other gauge anomalies, i.e., 
$\mathcal{A}_3\left[SU(3)_C^2 \times U(1)_{B-L}\right]$ and $\mathcal{A}_4\left[SU(2)_L^2 \times U(1)_{B-L}\right]$  
 trivially cancel).  These right-handed neutrinos can 
generate light neutrino masses via the type-I seesaw mechanism \cite{Minkowski:1977sc, Mohapatra:1979ia, Schechter:1980gr, GellMann:1980vs} and account for matter-antimatter asymmetry of the universe. However, we present below few other possible solutions to overcome these anomalies.
%

\subsection{Anomaly cancellation with additional fermions having exotic B-L charges}
In order to build an anomaly free $B-L$ gauge extended framework, the charges of the additional fermion content have to satisfy two simple equations given as \cite{Montero:2007cd}
\begin{equation}
\sum_{i=1}^{n_R} x_i^3 = 3 \quad {\rm{and}} \quad  \sum_{i=1}^{n_R} x_i = 3,
\label{anomalycon}
\end{equation}
where $n_R$ denotes the number of additional fermions and $x_i$ denotes the $B-L$ charge of each fermion. $n_R=1$ gives no solution and $n_R=2$ gives a complex solution. $n_R\geq 3$ is always suitable to have real solutions. For instance, choosing the charges as $-4,-4$ and $+5$ is one such solution satisfying \ref{anomalycon} and has been explored in \cite{Ma:2014qra,Ma:2015raa}. We show below the explicit check 
\begin{eqnarray}
&&\mathcal{A}_1\left[U(1)^3_{B-L} \right]= \mathcal{A}^{\rm SM}_1\left(U(1)^3_{B-L} \right) + \mathcal{A}^{\rm New}_1\left(U(1)^3_{B-L} \right)  =-3 + (4^3  +4^3  +(-5)^3)=0\, ,\nonumber \\
&&\mathcal{A}_2\left[\mbox{gravity}^2 \times U(1)_{B-L} \right]\propto   
     \mathcal{A}^{\rm SM}_2\left(U(1)_{B-L} \right) + \mathcal{A}^{\rm New}_2\left(U(1)_{B-L} \right)    =-3 + (4 +4+(-5))=0\, . \nonumber 
\end{eqnarray}

There could also be a different solution to cancel the gauge anomalies, where  one requires four additional fermions carrying fractional $B-L$ charges 
(first proposed in Ref.~\cite{Patra:2016ofq}). We briefly describe below,   how the  non-trivial  gauge anomalies   $\mathcal{A}_1\left(U(1)^3_{B-L} \right)$ 
and $\mathcal{A}_2\left(\mbox{gravity}^2 \times U(1)_{B-L} \right)$ 
 get cancelled by introducing four exotic fermions with fractional $B-L$ charges, i.e., $\xi_L (4/3)$,   $\eta_L(1/3)$,   $\chi_{1R}(-2/3)$ and   $\chi_{2R}(-2/3)$, where the corresponding $B-L$ charges are shown in the parenthesis, 
\begin{eqnarray}
&&\mathcal{A}_1\left[U(1)^3_{B-L} \right]= \mathcal{A}^{\rm SM}_1\left(U(1)^3_{B-L} \right) + \mathcal{A}^{\rm New}_1\left(U(1)^3_{B-L} \right) \nonumber \\
&&\hspace*{2cm}  =-3 +\left [\left(\frac{4}{3}\right) ^3 
    +\left(\frac{1}{3}\right) ^3 
    +\left(\frac{2}{3}\right) ^3
    +\left(\frac{2}{3}\right) ^3 \right ]=0\, ,\nonumber \\
&&\mathcal{A}_2\left[\mbox{gravity}^2 \times U(1)_{B-L} \right]\propto   
     \mathcal{A}^{\rm SM}_2\left(U(1)_{B-L} \right) + \mathcal{A}^{\rm New}_2\left(U(1)_{B-L} \right) \nonumber \\
&&\hspace*{2cm}     =-3 + \left [\left(\frac{4}{3}\right) 
    +\left(\frac{1}{3}\right) 
    +\left(\frac{2}{3}\right) 
    +\left(\frac{2}{3}\right)\right ]=0\, . \nonumber 
\end{eqnarray}

In this work, we consider the first category of anomaly free model built up based on $U(1)_{B-L}$ extension of the standard model which includes three neutral exotic fermions $N_{iR}$ (where $i=1,2,3$), with the $B-L$ charges $-4,-4$ and $+5$.  We include two more scalar singlets $\phi_1$ and $\phi_8$ to provide Majorana masses for all the exotic fermions  and also to spontaneously break the $B-L$ gauge symmetry. 
We also introduce a scalar dark matter $\phi_{\rm DM}$, singlet under the SM gauge group but charged under $U(1)_{B-L}$. It doesn't get any VEV, it does flow in the loop to generate light neutrino mass as discussed in Section-V. Fermionic dark matter of the current model has been explored in \cite{Singirala:2017cch}.
The particle content 
of the present model is given in Table \ref{tab:New_BL_DM}.


\begin{table}[tb!]
\begin{center}
\begin{tabular}{|c|c|c|c|c|}
	\hline
		& Field	& $SU(2)_L\times U(1)_Y$	& $U(1)_{B-L}$	\\
	\hline
	\hline
	Fermions& $Q_L \equiv(u, d)^T_L$		& $(\textbf{2}, 1/6)$	& $1/3$	\\
		& $u_R$					& $(\textbf{1}, 2/3)$	& $1/3$	\\
		& $d_R$					& $(\textbf{1}, -1/3)$	&$1/3$	\\
		& $\ell_L \equiv(\nu,~e)^T_L$		& $(\textbf{2}, -1/2)$	&  $-1$	\\
		& $e_R$					    & $(\textbf{1}, -1)$ &  $-1$	\\
        \hline
		& $N_{1R}$				    & $(\textbf{1}, 0)$	&   $-4$	\\
		& $N_{2R}$				    & $(\textbf{1}, 0)$	&   $-4$	\\
		& $N_{3R}$				    & $(\textbf{1}, 0)$	&   $5$	\\
	\hline
	Scalars	& $H$					& $(\textbf{2}, 1/2)$	&   $0$	\\
			& $\phi_{\rm DM}$			& $(\textbf{1}, 0)$	&   $n_{\rm DM}$	\\
			& $\phi_{1}$		& $(\textbf{1}, 0)$	&   $-1$	\\  
			& $\phi_{8}$			& $(\textbf{1}, 0)$	&   $8$		\\	
	\hline
	\hline
\end{tabular}
\caption{Particle spectrum and their charges of the proposed $U(1)_{B-L}$ model.}
\label{tab:New_BL_DM}
\end{center}
\end{table}

The relevant terms in the Lagrangian for fermions in the present model is given by 
\begin{eqnarray} 
&&\mathcal{L}^{\rm fermion}_{\rm Kin.}=
      \overline{Q}_L i \gamma^\mu \left(\partial_\mu+ig \frac{\vec{\tau}}{2} \cdot \vec{W}_\mu 
      +   \frac{1}{6} i\,g^\prime \,B_\mu + \frac{1}{3} i\,g_\text{BL} \,Z_\mu^\prime\right) Q_L   \nonumber \\
&&+  \overline{u_{R}} i \gamma^\mu \left(\partial_\mu
      +   \frac{2}{3} i \,g^\prime \,B_\mu + \frac{1}{3} i\,g_\text{BL} \,Z_\mu^\prime \right) u_{R} \nonumber \\
&&+  \overline{d_{R}} i \gamma^\mu \left(\partial_\mu
      -   \frac{1}{3} i\,g^\prime \,B_\mu + \frac{1}{3} i\,g_\text{BL} \,Z_\mu^\prime \right) d_{R} \nonumber \\
&&+  \overline{\ell_{L}} i \gamma^\mu \left(\partial_\mu+i g \frac{\vec{\tau}}{2} \cdot \vec{W}_\mu 
      -   \frac{1}{2} i\,g^\prime \,B_\mu - i\,g_\text{BL} \,Z_\mu^\prime \right) \ell_{L}            \nonumber \\
    &&+ \overline{e_{R}} i \gamma^\mu \left(\partial_\mu
      -  i\,g^\prime \,B_\mu -i\,g_\text{BL} \,Z_\mu^\prime  \right) e_{R}      \nonumber \\
&&+ \overline{N_{1R}} i \gamma^\mu \left(\partial_\mu
       -4 i\,g_\text{BL} \,Z_\mu^\prime \right) N_{1R} 
       + \overline{N_{2R}} i \gamma^\mu \left(\partial_\mu
       -4 i\,g_\text{BL} \,Z_\mu^\prime \right) N_{2R}\nonumber \\
&&+ \overline{N_{3R}} i \gamma^\mu \left(\partial_\mu
       +5 i\,g_\text{BL} \,Z_\mu^\prime \right) N_{3R}\;.
\end{eqnarray}
The interaction Lagrangian for the scalar sector is as follows
\begin{eqnarray}
\mathcal{L}^{\rm }_{\rm scalar} &=&
      \left(\mathcal{D}_\mu H \right)^\dagger \left(\mathcal{D}^\mu H\right) + \left(\mathcal{D}_\mu \phi_{\rm DM} \right)^\dagger \left(\mathcal{D}^\mu \phi_{\rm DM}\right)
      +\left(\mathcal{D}_\mu \phi_{1}\right)^\dagger \left(\mathcal{D}^\mu \phi_{1}\right)\nonumber \\
      &&
      +\left(\mathcal{D}_\mu \phi_{8}\right)^\dagger \left(\mathcal{D}^\mu \phi_{8}\right)
      -V\left( H,\phi_{\rm DM},\phi_1,\phi_8\right),
\end{eqnarray} 
where the covariant derivatives are
\begin{eqnarray} 
&&\mathcal{D}_\mu H = \partial_{\mu} H+i\,g \vec{W}_{\mu L}\cdot \frac{\vec{\tau}}{2}\, H  \,+\, i\frac{g^{\prime}}{2}B_{\mu} H\, , \nonumber \\
&&\mathcal{D}_\mu \phi_{\rm DM} =\partial_{\mu} \phi_{\rm DM} +  i~n_{\rm DM } g_{\rm BL} \,Z_\mu^\prime \phi_{\rm DM}  \, , \nonumber \\
&&\mathcal{D}_\mu \phi_{1} =\partial_{\mu} \phi_{1} - i g_{\rm BL} \,Z_\mu^\prime \phi_{1}  \, , \nonumber \\
&&\mathcal{D}_\mu \phi_{8} =\partial_{\mu} \phi_{8} + 8 i g_{\rm BL} \,Z_\mu^\prime \phi_{8}.
\end{eqnarray}
 The Yukawa interaction for the present model is given by
\begin{eqnarray}
\mathcal{L}_{\rm Yuk}&=&
     Y_u\, \overline{Q_{L}} \widetilde{H} u_{R} + Y_d \overline{Q_{L}}  H\, d_{R}
     +Y_e\, \overline{\ell_{L}} H e_{R} \nonumber \\
     &+&\sum_{\alpha=1,2} h_{\alpha 3} \phi_1 \overline{N^c_{\alpha R}} N_{3R}  + \sum_{\alpha, \beta=1,2} h_{\alpha \beta}\phi_8 \overline{N^c_{\alpha R}}   N_{\beta R}~, \label{dirac}
\end{eqnarray}
with $\widetilde{H}=i\sigma_2 H^*$. From the above Yukawa interaction terms, one can write the exotic fermion mass matrix and diagonalize it to obtain non-zero masses to all the Majorana mass eigenstates.

\subsection{Scalar Potential Minimization and Stability criteria}
The scalar potential of this model is given by
 \begin{align}
V(H,\phi_{\rm DM},\phi_1,\phi_8)= & \mu^2_{\rm H}  H^\dagger H + \lambda_{\rm H} (H^\dagger H)^2 + \mu^2_1 \phi^\dagger_1 \phi_1 + \lambda_1 (\phi^\dagger_1 \phi_1)^2 + \mu^2_8 \phi^\dagger_8 \phi_8  + \lambda_8 (\phi^\dagger_8 \phi_8)^2 \nonumber \\
     + & \mu^2_{\rm DM} \phi_{\rm DM}^\dagger {\phi_{\rm DM}} + \lambda_{\rm DM} (\phi_{\rm DM}^\dagger {\phi_{\rm DM}})^2  +\lambda_{\rm H1} (H^\dagger H) (\phi^\dagger_1 \phi_1)
      +\lambda_{\rm H8} (H^\dagger H) (\phi^\dagger_8 \phi_8)\nonumber \\
      +& \lambda_{18} (\phi^\dagger_1 \phi_1) (\phi^\dagger_8 \phi_8) + \lambda_{\rm HD} (H^\dagger H) (\phi_{\rm DM}^\dagger \phi_{\rm DM}) + \lambda_{\rm D1} (\phi_{\rm DM}^\dagger \phi_{\rm DM}) (\phi^\dagger_1 \phi_1) \nonumber \\
      +& \lambda_{\rm D8} (\phi_{\rm DM}^\dagger \phi_{\rm DM}) (\phi^\dagger_8 \phi_8),
      \label{scalar}
\end{align} 
where $\phi_{\rm DM} = \frac{S+iA}{\sqrt{2}}$, the scalar  fields $H=(H^+,H^0)^T$, $\phi_1$ and $\phi_8$ can be parametrized in terms of real scalars and pseudo scalars as
\begin{align}
&H^0 =\frac{1}{\sqrt{2} }(v+h)+  \frac{i}{\sqrt{2} } A^0\,, \nonumber \\
& \phi^0_1 = \frac{1}{\sqrt{2} }(v_1+h_1)+  \frac{i}{\sqrt{2} } A_1\,, \nonumber \\
& \phi^0_8 = \frac{1}{\sqrt{2} }(v_8+h_8)+  \frac{i}{\sqrt{2} } A_8\, ,
\end{align}
where $\langle H\rangle=(0,v/\sqrt2)^T$, $\langle \phi_1\rangle=v_1/\sqrt2$, and $\langle \phi_8\rangle=v_8/\sqrt2$. 
\subsection{Vacuum stability criteria and Unitarity constraints}
The vacuum stability conditions of the scalar potential are given by \cite{Kannike:2012pe, Kannike:2016fmd}
\bea
&&\lambda_{\rm H} \ge 0,~ \lambda_{\rm HD} \ge 0,~ \lambda_{\rm DM} \ge 0,~ \lambda_1 \ge 0, ~ \lambda_8 \ge 0,\nn\\
&&\lambda_{\rm D1} + \sqrt{\lambda_{\rm DM} \lambda_1} \ge 0,~ \lambda_{\rm D8} + \sqrt{\lambda_{\rm DM} \lambda_8} \ge 0,~
\lambda_{18} + \sqrt{\lambda_1 \lambda_8} \ge 0, \nn\\
&&\sqrt{\lambda_{\rm DM} \lambda_1 \lambda_8} + \lambda_{\rm D1} \sqrt{\lambda_8} + \lambda_{\rm D8} \sqrt{\lambda_1} + \lambda_{18} \sqrt{\lambda_{\rm DM}} \ge 0.
\eea
Now we apply the tree-level perturbative unitarity constraints on the scattering processes of the scalar sector. The formula for the zeroth partial wave amplitude \cite{Lee:1977eg} is  given by
\begin{equation}
a_0=\frac{1}{32\pi} \sqrt{\frac{4 p^{\rm CM}_f p^{\rm  CM}_i}{s}}\int_{-1}^{+1}  d (\cos\theta) ~T_{2 \to 2} .
\end{equation}
Here  $p_{i,(f)}^{\rm CM}$ is the the centre of mass (CoM) momentum of the initial (final) state, $s$ is the 
(CoM) energy,
and   $T_{2 \to 2}$ denotes the full amplitude of each $2\to2$ scattering processes. At high energies, the partial wave amplitudes i.e., the quartic couplings gets constrained from perturbative unitarity requirement $\left| {\rm Re} (a_0)\right| \le \frac{1}{2}$, giving
\bea
&&\lambda_{\rm H}, \lambda_{1}, \lambda_{8}, \lambda_{\rm DM} \le \frac{4\pi}{3},\nn\\
&&\lambda_{\rm HD}, \lambda_{\rm D1}, \lambda_{\rm D8}, \lambda_{\rm H1},\lambda_{\rm H8},\lambda_{\rm 18} \le 4\pi. 
\eea
\subsection{Mixing in scalar sector}
In the scalar sector, the CP-even scalar mass matrix takes the form 
\begin{align}
	M_{E}^2
	=
	\begin{pmatrix}
		2\lambda_H v^2	& \lambda_{H1}vv_1& \lambda_{H8}vv_8	\\
		\lambda_{H1}vv_1	&2\lambda_1 v_1^2	& \lambda_{18}v_1v_8	\\
		\lambda_{H8}vv_8	&\lambda_{18}v_1v_8	& 2\lambda_8 v_8^2				\\
	\end{pmatrix}. 
\end{align}
We assume that the Higgs doublet $H$ mixes equally with the two singlets and the mixing is small so that the decay width of Higgs is consistent with LHC limits. We also consider the VEVs of the new singlets $v_1 \simeq v_8 \gg v$ and the couplings $\lambda_{H1,H8}\ll \lambda_H$, $\lambda_1 \simeq \lambda_8$ then the mass matrix takes a simple form
\begin{align}
	M_E^2
	\simeq
	\begin{pmatrix}
		~a~	& ~a~ & ~a~	\\
		~a~	& ~y~	& ~b~	\\
		~a~	& ~b~	& ~y~ \\
	\end{pmatrix}. 
	\label{massMatrix}
\end{align}
Under the assumption of minimal Higgs mixing, the unitary matrix that connects the flavor and mass states is
\begin{align}
U \simeq
\begin{pmatrix}
 1 & ~\beta  \cos \alpha -\beta  \sin \alpha  & ~\beta  \cos \alpha +\beta  \sin \alpha  \\
 -\beta  &~ \cos \alpha  & ~\sin \alpha   \\
 -\beta  &~ -\sin \alpha  &~ \cos \alpha  \\
\end{pmatrix}.
\end{align}
Here $\beta = \frac{a}{b+y-a}$ is the mixing parameter for $H-\phi_{1,8}$ and  $\alpha = \frac{5\pi}{4}$ denotes $\phi_1-\phi_8$ mixing, obtained from the normalized eigenvector matrix  of $M_E^2$ (\ref{massMatrix}). Thus, the relation between flavor and mass eigenstates is given by
\begin{align}
\begin{pmatrix}
 h \\
 h_1 \\
 h_8 \\
 \end{pmatrix} = U \begin{pmatrix} H_1\\ H_2 \\ H_3 \end{pmatrix}=
\begin{pmatrix}
 H_1-H_3 \beta  \sqrt{2} \\
 -H_1 \beta - \frac{H_2}{\sqrt{2}} - \frac{H_3}{\sqrt{2}} \\
 -H_1 \beta + \frac{H_2}{\sqrt{2}} - \frac{H_3}{\sqrt{2}} \\
 \end{pmatrix}.
\end{align}
The scalar couplings can be written as
\begin{eqnarray}
&&2\lambda_{H}v^2 = \lambda_{H1}vv_1 =  \frac{M_{H_1}^2}{(1-2\beta+2\beta^2)},\nn\\
&&2\lambda_{1}v_1^2 = 2\lambda_{8}v_8^2 = \frac{(\beta+1)M_{H_3}^2 + (1+\beta+4\beta^2)M_{H_2}^2}{2(1+\beta+4\beta^2)},\nn\\
&&\lambda_{18}v_1v_8  = \frac{(\beta+1)M_{H_3}^2 - (1+\beta+4\beta^2)M_{H_2}^2}{2(1+\beta+4\beta^2)}.
\label{scouplings}
\end{eqnarray}
Here $H_1$ denotes the SM Higgs with $M_{H_1} = 125.09$ GeV with $v = 246$ GeV. The mixing angle $\beta$ can be written in terms of the physical scalar masses as
\begin{equation}
\beta=\frac{-M_{H_1}^2+M_{H_3}^2-\sqrt{-15 M_{H_1}^4-10 M_{H_3}^2 M_{H_1}^2+M_{H_3}^4}}{4 \left(2 M_{H_1}^2+M_{H_3}^2\right)}.
\end{equation}
Since the Higgs mass ($M_{H_1}$) is fixed, the mass parameter $M_{H_3}$ defines the amount of mixing i.e., say $M_{H_3} \ge 1$ TeV implies $\beta \le 0.016$.
Moving to CP-odd components, the linear combination (denoted by $A_{\rm G}$) of $A_1$ and $A_8$ is eaten up by $Z^{\prime}$ and the other orthogonal combination, $A_{\rm NG}$ remains as a massless physical Goldstone (NG), which are given by
\begin{eqnarray}
&&A_{\rm G} = - \frac{8v_8}{\sqrt{v_1^2 + 64v_8^2}}A_8 + \frac{v_1}{\sqrt{v_1^2 + 64v_8^2}}A_1 ,\nn\\
&&A_{\rm NG} = \frac{v_1}{\sqrt{v_1^2 + 64v_8^2}}A_8 + \frac{8v_8}{\sqrt{v_1^2 + 64v_8^2}}A_1.
\end{eqnarray}
As per the assumption $v_1 \simeq v_8$, one can see that $A_{\rm G}$ gets major contribution from $A_8$ and $A_{\rm NG}$ is maximally composed of $A_1$. This massless mode can couple to the new fermion  and scalar sectors as shown in Appendix \ref{apdx}. In SM, it can only couple to Higgs as we considered non-zero mixing between $H$ and the new scalar singlets.
It can give rise to an additional decay channel contributing to the invisible width of SM Higgs, given as
\begin{equation}
\Gamma(H_1\to A_{\rm NG}A_{\rm NG}) \simeq \frac{M_{H_1}^3 \sin^2\beta}{32\pi}\left(\frac{v_1^3 + 64v_8^3}{v_1v_8(v_1^2+64v_8^2)}\right)^2~,
\end{equation}
where $\beta$ denotes the mixing between $H$ and $\phi_1$. 
The invisible branching ratio of Higgs is given as
\begin{equation}
\text{Br}_{\rm inv} = \frac{\Gamma(H_1\to A_{\rm NG}A_{\rm NG})}{\Gamma(H_1\to A_{\rm NG}A_{\rm NG}) + \cos^2\beta ~\Gamma^{\rm Higgs}_{\rm SM}}.
\end{equation}
Using the constraint, $\text{Br}_{\rm inv}\simeq 20\%$ \cite{Belanger:2013kya,Giardino:2013bma}, $\Gamma^{\rm Higgs}_{\rm SM} \simeq 4$ MeV, we obtain the upper limit on the mixing angle as
\begin{equation}
|{\tan\beta}| \lesssim 2.2\times 10^{-4}\times \left(\frac{v_1}{\text{GeV}}\right).
\end{equation}
Moreover, if the NG stays in thermal equilibrium with ordinary matter until muon annihilation, then it mimics as fractional cosmic neutrinos contributing nearly 0.39 to the effective number of neutrino species \cite{Weinberg:2013kea} to give $N_{\rm eff} = 3.36^{+0.68}_{-0.64}$ at $95\%$ C.L., a remarkable agreement with Planck data. 
This illustration was done by working in the low mass regime of the physical scalar ($\simeq 500$ MeV).
However, in the present work, we consider higher mass regime for the physical scalar spectrum and discuss the effect of NG on relic density in the upcoming section.
\subsection{Stability of singlet scalar dark matter}
\begin{figure}[h!]
\centering
\includegraphics[width=0.5\linewidth]{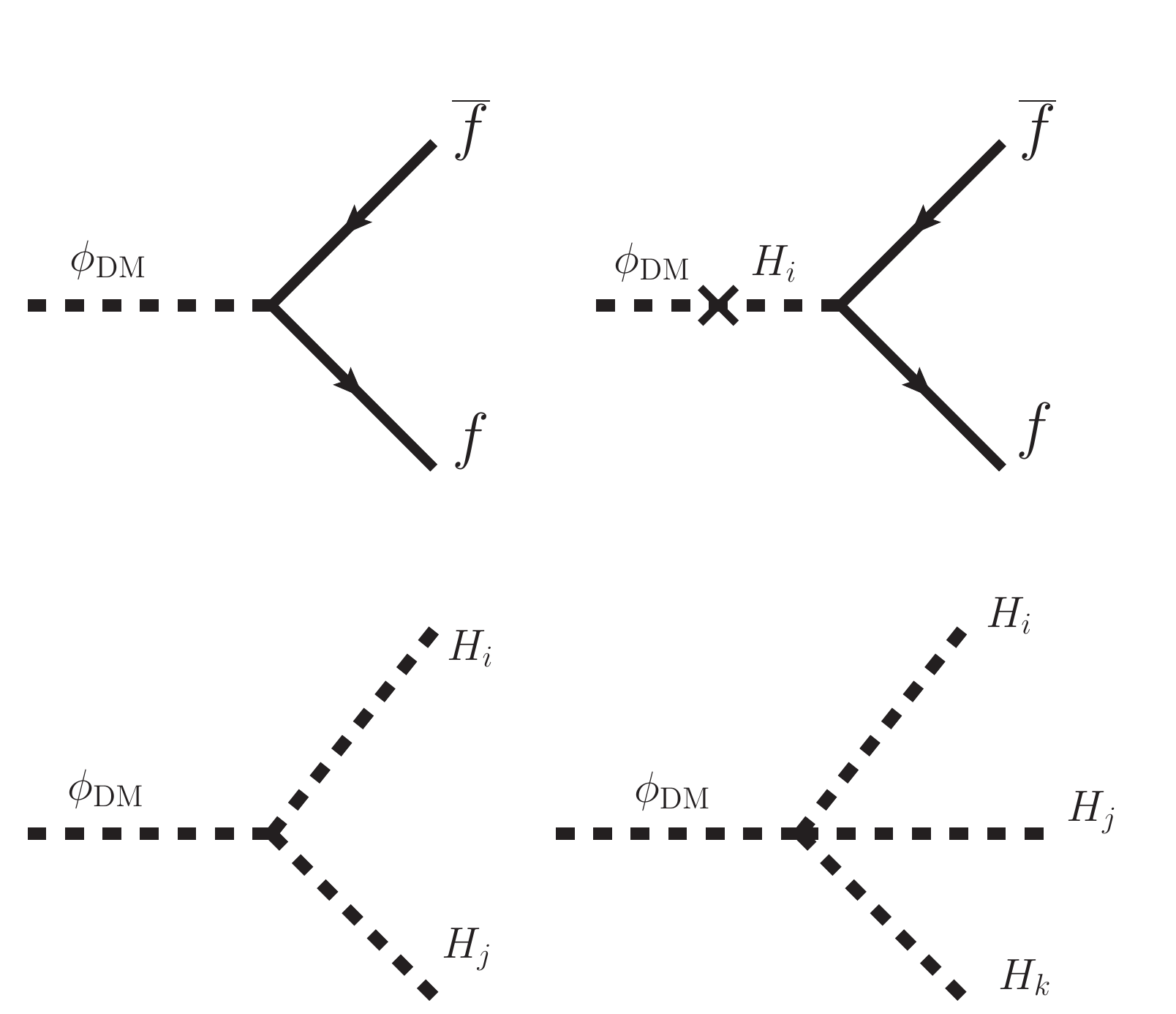}
\caption{Feynman diagrams leading to decay of scalar singlet dark matter $\phi_{\rm DM}$. The choice of $B-L$ 
charge to forbid these decay and stability of scalar singlet dark matter $\phi_{\rm DM}$ is discussed in the text. }
\label{decayDM}
\end{figure}

Dark matter particle has to be electrically neutral and should be stable over cosmological time scales. With this motivation numerous frameworks 
were proposed based on an unbroken discrete symmetry \cite{Ma:2006km, Belanger:2014bga} forbidding the decay of DM. Furthermore, this discrete symmetry is expected to break at Planck scale and thus, induce the decay of DM making it unstable. In the present model, we don't assume any ad-hoc discrete symmetry as such which can stabilize the DM.  Rather we choose the $B-L$ 
charge (say  $n_{\rm DM}$) in such a way that there won't be any decay channel as displayed in Fig. \ref{decayDM} for the DM   $\phi_{\rm DM}$ \cite{Iso:2009nw}. For example, to avoid the cubic term in the scalar potential of the form  
$\phi_{\rm DM} H_i H_j$ where $H_i, H_j$ denote the physical masses for any of the scalars $H, \phi_1$ or $\phi_8$, the possible values of  $n_{\rm DM} =  0, \pm 2, \pm 7, \pm 9, \pm 16$ are not allowed.  
Similarly if we don't want term like $\phi_{\rm DM} H_i H_j H_k$, the value of $n_{\rm DM}$ is restricted to $ n_{\rm DM} \neq \pm 1, \pm 3, \pm 6, \pm 8, \pm 10$. 
Thus, the allowed values of $n_{\rm DM}$ are $\pm 4, \pm 5$ and fractional charges. The approach of ensuring stability of scalar DM particle  with the model structure has been recently implemented in a $B-L$ model with right-handed neutrinos \citep{Rodejohann:2015lca}, while our model is one such variant with a modified scalar content and variety of exotic charges assigned to the additional fermion content of the $B-L$ model.\\

We choose $n_{\rm DM} = 4$ to ensure the stability of the scalar singlet $\phi_{\rm DM}$ and study its phenomenology in the prospects of dark matter observables such as relic abundance and direct detection cross section. Based on the structure of the model built, the DM can have scalar and gauge portal interactions. We proceed to study in detail the behaviour of DM observables separately in dual portal scenario.\\

\section{$Z^{\prime}$ portal phenomenology}
\subsection{Relic abundance}
\begin{figure}[t!]
\begin{center}
\includegraphics[width=0.3\linewidth]{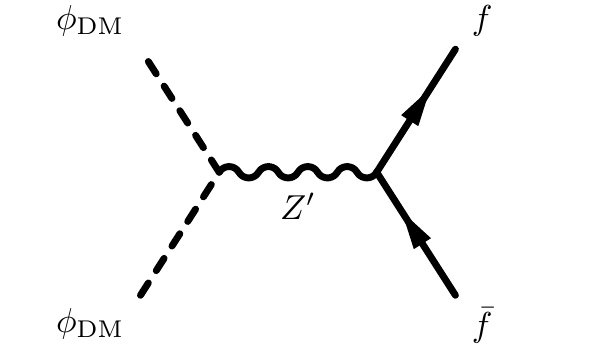}
\vspace{0.01 cm}
\includegraphics[width=0.3\linewidth]{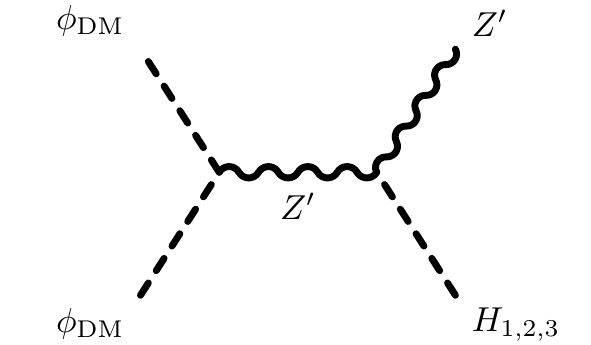}
\vspace{0.01 cm}
\includegraphics[width=0.3\linewidth]{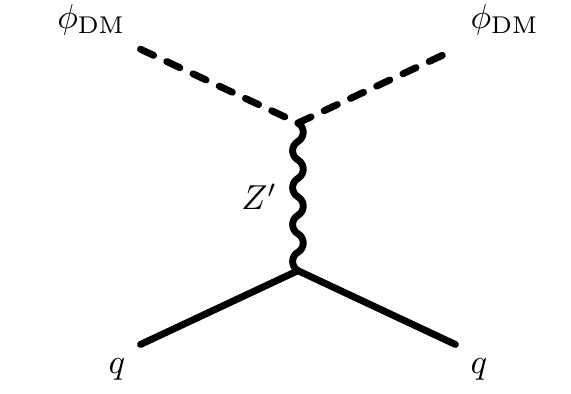}
\caption{Feynman diagrams for dark matter annihilation (left and middle) and scattering of DM from nucleon/quark (right panel) through $Z'$ exchange. First two diagrams contribute to the relic density observable and the third one is appropriate in direct detection studies.}
\label{zpfeyn}
\end{center}
\end{figure}

The channels that contribute to relic density are shown in the left and middle panels of Fig. \ref{zpfeyn} and the  expression for the corresponding annihilation cross sections are
\begin{eqnarray}
\hat{\sigma}_{ff} &=& \sum_f \frac{n^2_{\rm DM} (n^f_{\rm BL})^2g^4_{\rm BL} c_f}  {12 \pi s } \frac{(s-4 M^2_{\rm DM})(s + 2M^2_f)}{\left[(s-M^2_{Z^{\prime}})^2 + M^2_{Z^{\prime}} \Gamma^2_{Z^{\prime}}\right]} \frac{(s-4 M^2_f)^{\frac{1}{2}}}{(s-4 M^2_{\rm DM})^{\frac{1}{2}}},\nn\\
\hat{\sigma}_{Z^{\prime} H_i} &=&  \frac{n^2_{\rm DM} g^6_{\rm BL} (C_{H_i})^2}{16 \pi s }\frac{(s-4 M^2_{\rm DM})}{\left[(s-M^2_{Z^{\prime}})^2 + M^2_{Z^{\prime}} \Gamma^2_{Z^{\prime}}\right]} \left[1 + \frac{(s-(M_{Z^\prime}+M_{H_i})^2)(s-(M_{Z^\prime}-M_{H_i})^2)}{12s M^2_{Z^\prime} } \right]\nn\\
&&\frac{\Big [(s-(M_{Z^\prime}+M_{H_i})^2)(s-(M_{Z^\prime}-M_{H_i})^2) \Big ]^\frac{1}{2}}{\left [s(s-4 M^2_{\rm DM}) \right ]^{\frac{1}{2}}}, 
\label{zpann}
\end{eqnarray}
where $i = 1, 2,3$ and
\begin{align*}
    C_{H_1} &= 2\beta(64v_8-v_1),\\
    C_{H_2} &= \sqrt{2}(64v_8-v_1),\\  
C_{H_3} &= \sqrt{2}(64v_8+v_1).	
	\label{}
\end{align*}

The parameters $c_f$ and $n^f_{\rm BL}$ denote the color charge and the $B-L$ charge of the fermion $f$ with mass $M_f$. $M_{Z^{\prime}}$ is the mass of the heavy gauge boson $Z^{\prime}$ given by $M_{Z^{\prime}} = g_{\rm BL} \sqrt{v_1^2 + 64 v_8^2 }$ with the decay width $\Gamma_{Z^{\prime}}$. The relic abundance of dark matter is computed by
\begin{equation}
\label{eq:relicdensity}
\Omega \text{h}^2 = \frac{2.14 \times 10^{9} ~{\rm{GeV}}^{-1}}{ {g_\ast}^{1/2} M_{\rm{pl}}}\frac{1}{J(x_f)},
\end{equation}
where $M_{\rm{pl}}=1.22 \times 10^{19} ~\rm{GeV}$ is the Planck mass, $g_\ast = 106.75$ denoting the total number of effective relativistic degrees of freedom, 
and $J(x_f)$ reads as
\begin{equation}
J(x_f)=\int_{x_f}^{\infty} \frac{ \langle \sigma v \rangle (x)}{x^2} dx.
\end{equation}
The thermally averaged annihilation cross section $\langle \sigma v \rangle$ is given by
\begin{equation}
 \langle\sigma v\rangle (x) = \frac{x}{8 M_{\rm DM}^5 K_2^2(x)} \int_{4 M_{\rm DM}^2}^\infty \hat{\sigma} \times ( s - 4 M_{\rm DM}^2) \ \sqrt{s} \ K_1 \left(\frac{x \sqrt{s}}{M_{\rm DM}}\right) ds.
\end{equation}
The functions $K_1$, $K_2$ denote the modified Bessel functions and $x = M_{\rm DM}/T$, where $T$ is the temperature. The analytical expression for the freeze out parameter $x_f$ is given as
\begin{equation}
x_f= \ln \left( \frac{0.038 \ g \ M_\text{Pl} \ M_{\rm DM} \ \langle\sigma v\rangle (x_f) }{({g_\ast x_f})^{1/2}} \right).
\end{equation}
Here $g$ is the count of number of degrees of freedom of the dark matter particle $S$.
\begin{figure}[htb]
\begin{center}
\includegraphics[width=0.48\linewidth]{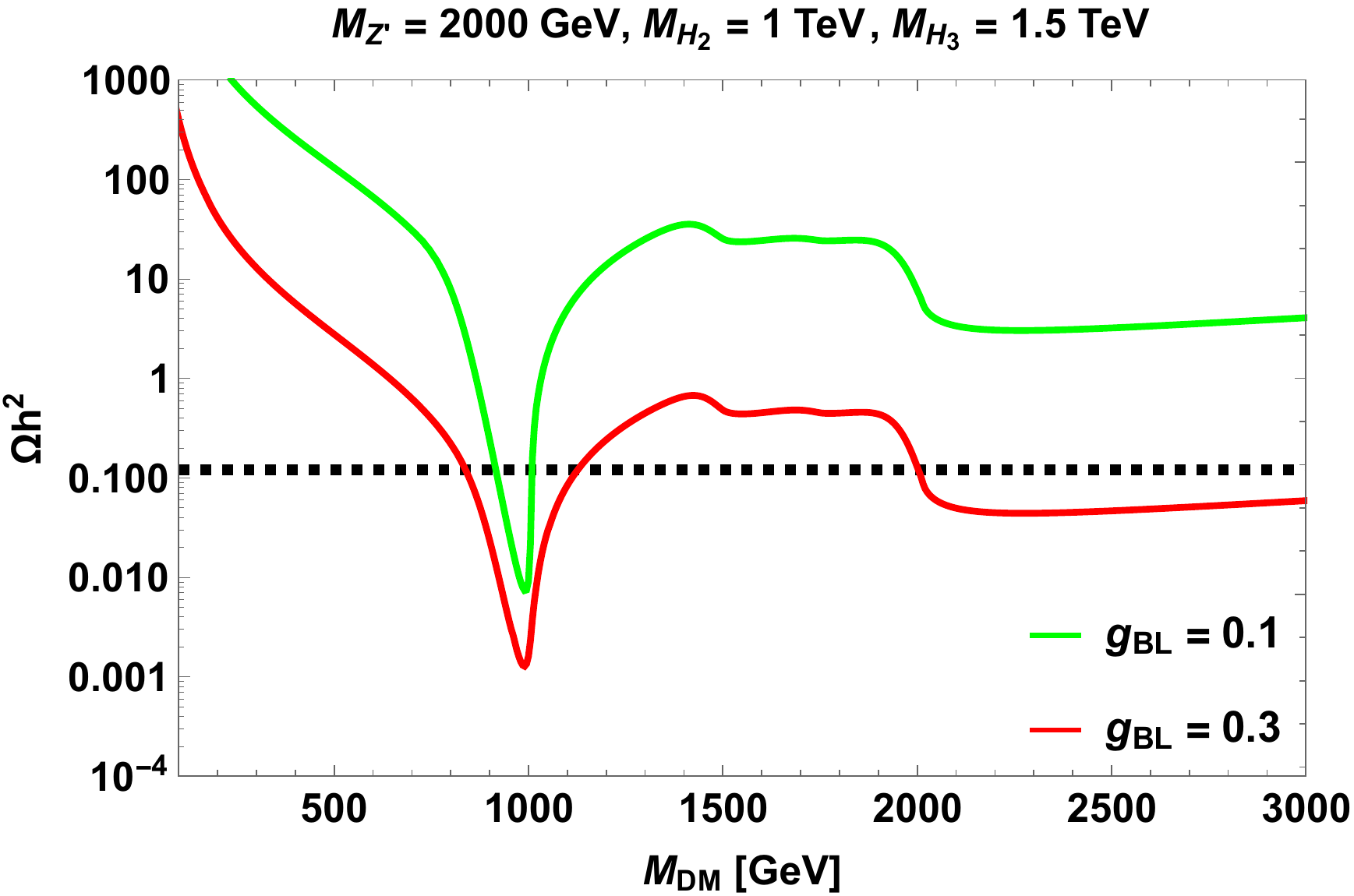}
\vspace{0.2 cm}
\includegraphics[width=0.48\linewidth]{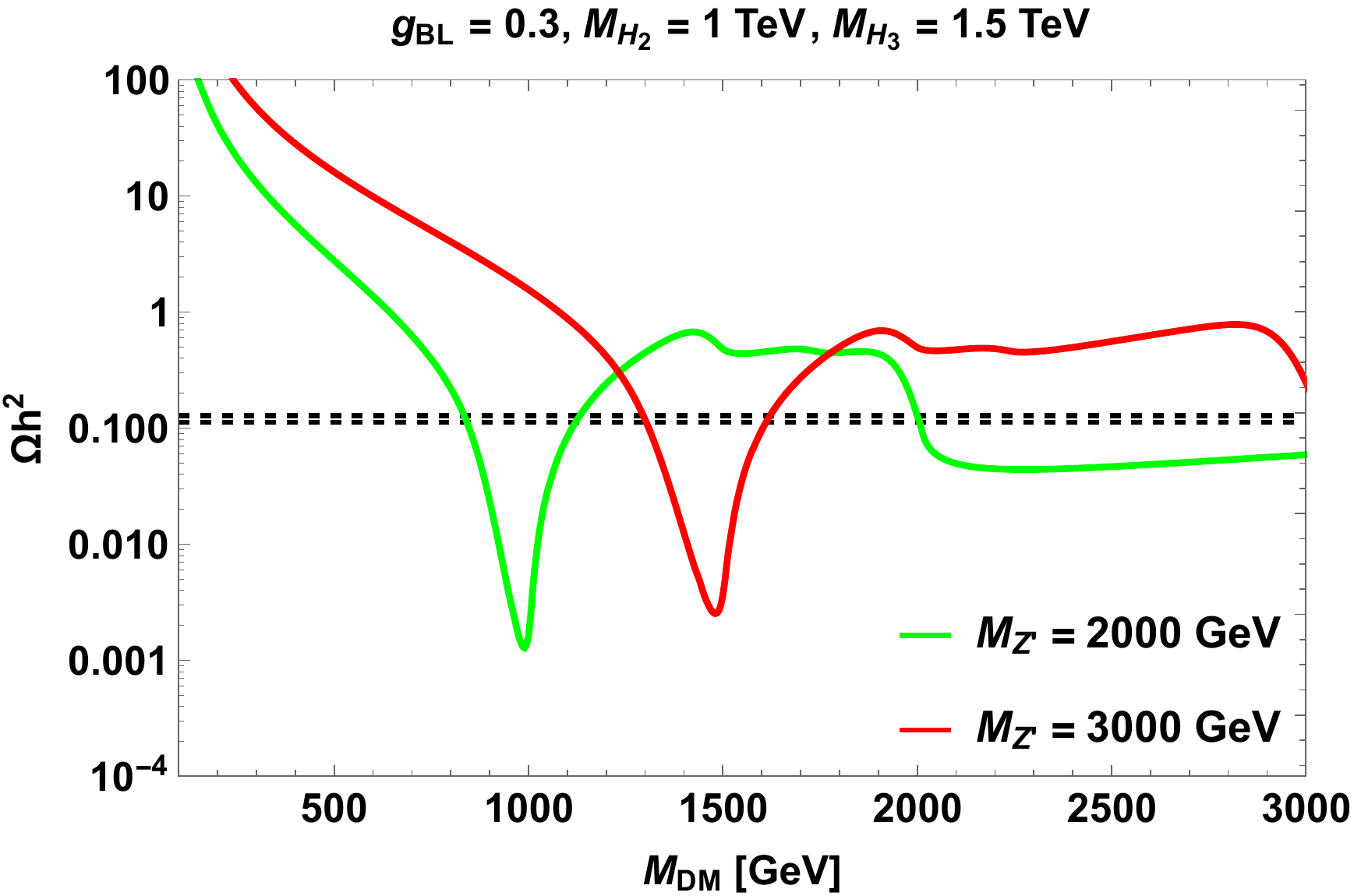}
\caption{Variation of relic abundance $\Omega \text{h}^2$ with the mass of DM shown  with two representative values of  $g_{\rm BL}$ (left panel) and  $M_{Z^{\prime}}$    (right panel) for $n_{\rm DM} = 4$.  Here the horizontal dashed lines denote the $3 \sigma$ range in current relic density \citep{Ade:2015xua}.}
\label{ZpOmega}
\end{center}
\end{figure}
We have implemented the model in LanHEP \citep{Semenov:1996es} to produce the model files required for micrOMEGAs \cite{Pukhov:1999gg, Belanger:2006is, Belanger:2008sj} package to compute the relic abundance of scalar DM. 
\begin{table}[t!]
\begin{center}
\begin{tabular}{|c|c|c|c|c|c|c|c|c|c|}
	\hline
	Parameters & $n_{\rm DM}$	&$M_{H_{2}}$ [GeV]	& $M_{H_{3}}$ [GeV] & $v_{1,8}$   [GeV] & $\beta$\\
	\hline
	Values &$4$ & $1000$ & $1500$ & $2000$  & $0.007$\\
	\hline
\end{tabular}
\caption{Fixed parameters for $Z^{\prime}$-mediated DM observables.}
\label{fixed}
\end{center}
\end{table}
The parameters that are fixed during the analysis are shown in Table. \ref{fixed}. The flexibility of gauge portal study is that, just two parameters are relevant i.e., $g_{\rm BL}$ and $M_{Z^{\prime}}$. The value of $n_{\rm DM}$ not only stabilizes the DM paricle but also scales the annihilation cross section thereby showing up in relic density. Fig. \ref{ZpOmega} displays the variation of DM abundance $\Omega \text{h}^2$ with the singlet DM mass $M_{\rm DM}$ and the behavior with various  parameters.  All the curves in Fig. \ref{ZpOmega} reach the current relic density of PLANCK \cite{Ade:2015xua} near the resonance ($M_{\rm DM} \simeq \frac{M_{Z^{\prime}}}{2}$). The gauge coupling $g_{\rm BL}$ scales the annihilation cross section i.e., lower couplings give lower annihilation cross section. The channels $SS \to f\bar{f}$ drive the relic density until the channels $SS \to Z^{\prime} H_1$, $SS \to Z^{\prime} H_2$ and $SS \to Z^{\prime} H_3$ get kinematically allowed.
\subsection{Direct searches}
\begin{figure}[htb]
\begin{center}
\includegraphics[width=7.0 cm,height= 5.0 cm, clip]{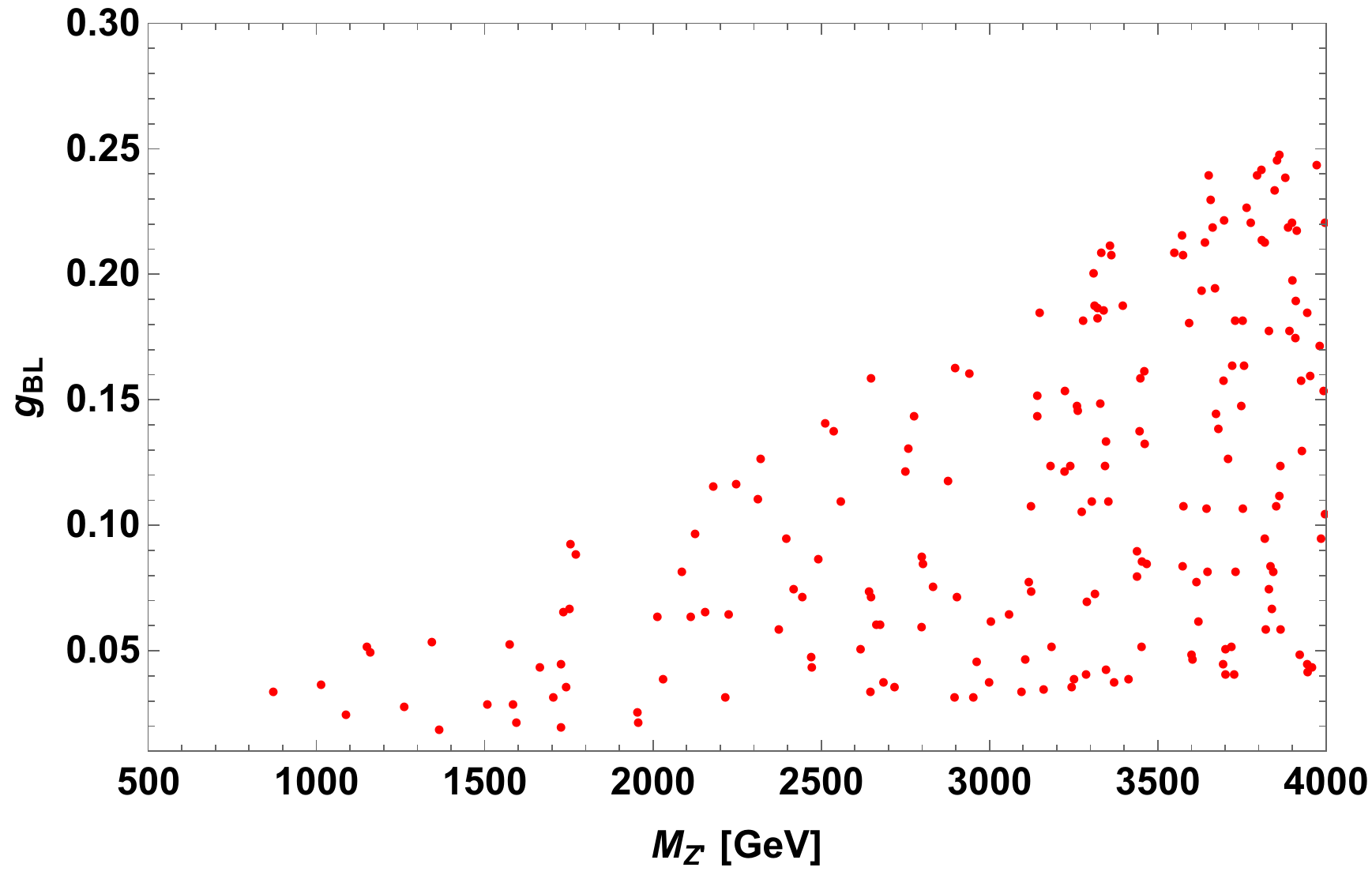}
\vspace{0.2 cm}
\includegraphics[width=7.0 cm,height= 5.0 cm, clip]{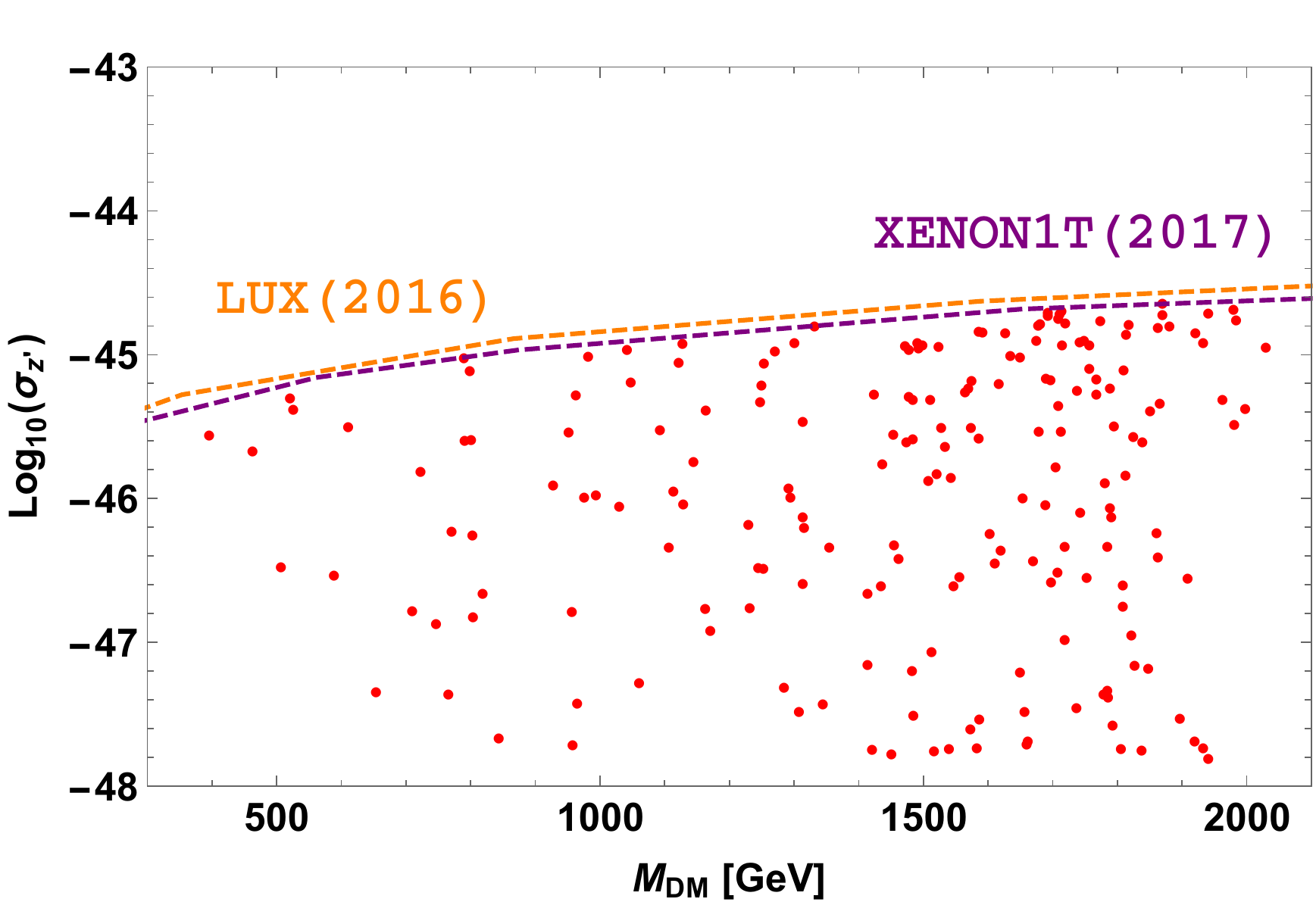}
\caption{Left panel denotes the parameters space in the plane of $(M_{Z^{\prime}}, g_{\rm BL})$ that satisfy the current relic density \cite{Ade:2015xua} in $3\sigma$ range and XENON1T \cite{Aprile:2017iyp}. The right panel depicts the WIMP-nucleon SI cross section with the mass of the scalar DM for the parameter space shown in the left panel. Here, the horizontal dashed lines denote the current bounds on spin-independent WIMP-nucleon cross section from the direct detection experiments LUX \cite{Akerib:2016vxi} and XENON1T  \cite{Aprile:2017iyp}.}
\label{Zpdirect}
\end{center}
\end{figure}

Now we look for the constraints on the model parameters due to direct detection limits. 
The interaction terms for $Z^{\prime}$-mediated t-channel processes shown in the extreme right panel of Fig. \ref{zpfeyn} is  given as
\bea \mathcal{L}&\supset&
-{n_{\rm DM}}ig_{\rm BL}Z^{\prime}_{\mu}\left(S\partial^{\mu}A-A\partial^{\mu}S  \right)
-\frac{1}{3}g_{\rm BL}Z^{\prime}_{\mu}\bar{u}\gamma^\mu u
-\frac{1}{3}g_{\rm BL}Z^{\prime}_{\mu}\bar{d}\gamma^\mu d. \eea
Thus, the effective Lagrangian follows as
\bea%
i\mathcal{L}_{\mathrm{eff}}\supset
-\frac{{n_{\rm DM}} g_{\rm BL}^2}{3 M_{Z^{\prime}}^2}
\left(S\partial^{\mu}A-A\partial^{\mu}S  \right)\bar{u}\gamma_\mu u
-\frac{{n_{\rm DM}}g_{\rm BL}^2}{3 M_{Z^{\prime}}^2}
\left(S\partial^{\mu}A-A\partial^{\mu}S  \right) \bar{d}\gamma_\mu d.%
\eea\label{eff1} %
The DM-nuclei cross-section of the singlet scalar DM mediated by the gauge boson $Z^{\prime}$ is given by \cite{Goodman:1984dc, Jungman:1995df, Khalil:2011tb, Chiang:2012qa,Kanemura:2011mw,Zheng:2010js,Farzan:2012sa,Kohri:2013sva} 
\be%
\sigma^{\rm N}_{\rm SI} = \frac{1}{16\pi}\left(\frac{M_{\rm N} M_{\rm DM}}{M_{\rm N} + M_{\rm DM}}\right)^2 \left|b_{\rm N}\right|^2, %
\ee%
where $M_{\rm N}$ is the nuclei mass and the coefficient $b_{\rm N}$ is given by%
\bea %
b_{\rm N}=(A-Z)b_n+Zb_p, \quad b_n=b_u+2b_d, \quad b_p=2b_u+b_d, %
\eea %
Here $Z$ and $A$ denote the atomic and the mass number respectively. The parameters $b_u$ and $b_d$ of the effective Lagrangian are defined as
\begin{equation}
\mathcal{L}_{\mathrm{eff}}= b_q X^\mu \bar{q}\gamma^\mu q, \quad
{\rm{where}} \quad q=(u~,~d).
\end{equation}
In the present model, $X^\mu$ takes the form of the vector current given by  $X^\mu\simeq iS \partial^\mu A-iA\partial^\mu S$.
Thus, one can find the value of $b_{p,n}$ as
\be
b_p=b_n=i\frac{{n_{\rm DM}} g_{\rm BL}^2}{M_{Z^{\prime}}^2}. \nn%
\ee%
Therefore, $b_{\rm N}$ can have the value
\be%
b_{\rm N}=iA\frac{{n_{\rm DM}} g_{\rm BL}^2}{M_{Z^{\prime}}^2}. \nn%
\ee %
Thus, the DM-nuclei SI contribution is given by%
\be
\sigma^{\rm N}_{\rm SI}=\frac{1}{16\pi}\left(\frac{M_{\rm N} M_{\rm DM}}{M_{\rm N} + M_{\rm DM}}\right)^2|A|^2\frac{{n^2_{\rm DM}} g_{\rm BL}^4}{M_{Z^{\prime}}^4}.\nn
\ee
For single nucleon, the above expression becomes
\be
\sigma_{Z^{\prime}}=\frac{{\mu}^2}{16\pi}\frac{{n^2_{\rm DM}} g_{\rm BL}^4}{M_{Z^{\prime}}^4}.
\ee
where $\mu = \left(\frac{M_n M_{\rm DM}}{M_n + M_{\rm DM}}\right)$ is the reduced mass of DM-nucleon with $M_n$ being the nucleon mass. 
We see that the $B-L$ charge $n_{\rm DM}$ remains as a scaling parameter in $\sigma_{Z^{\prime}}$ alike relic density in Eqn. \ref{zpann}. It is convenient to write (in $\text{cm}^2$) as
\begin{equation}
 \sigma_{Z^{\prime}}  = 7.75 \times 10^{-42} \left( \frac{\mu}{1 ~{\rm{GeV}}}\right)^2 \times {n_{\rm DM}}^2 \times \left( \frac{1 ~{\rm{TeV}}}{\left(\frac{M_{Z^{\prime}}}{g_{\rm BL}}\right)}\right)^4 .\hspace{2.5 truecm}
\end{equation}
We show in the left panel Fig. \ref{Zpdirect}, the parameter space that satisfies the $3\sigma$  range in the current relic density limit \cite{Ade:2015xua} and the most stringent direct detection bound form XENON1T \cite{Aprile:2017iyp}. The right panel shows the WIMP-nucleon spin-independent cross section with the DM mass for the parameter space shown in the left panel.  

\subsection{Collider bounds}
\begin{figure}[htb]
\begin{center}
\includegraphics[width=8.0 cm,height= 5.0 cm, clip]{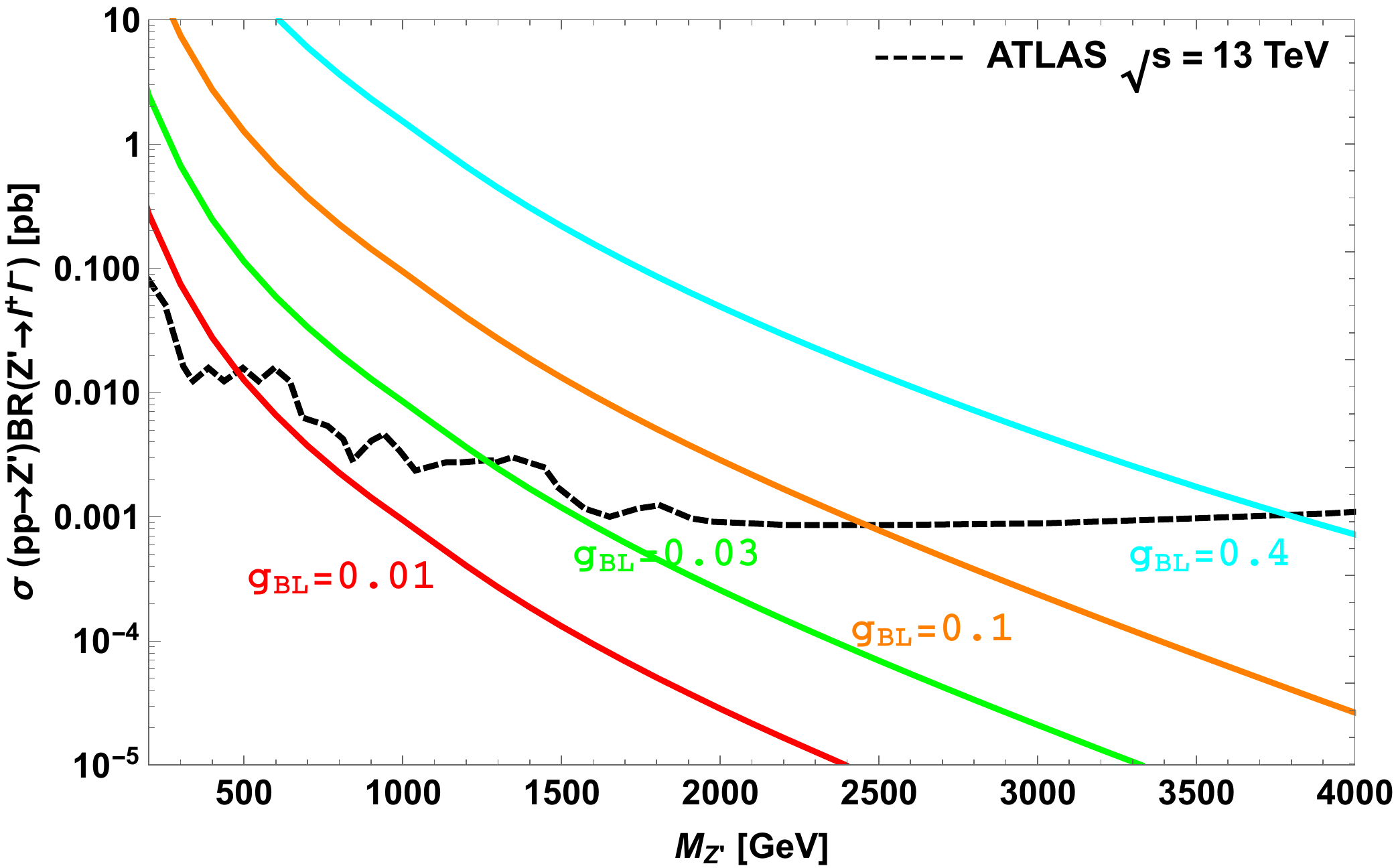}
\vspace{0.2 cm}
\includegraphics[width=8.0 cm,height= 5.0 cm, clip]{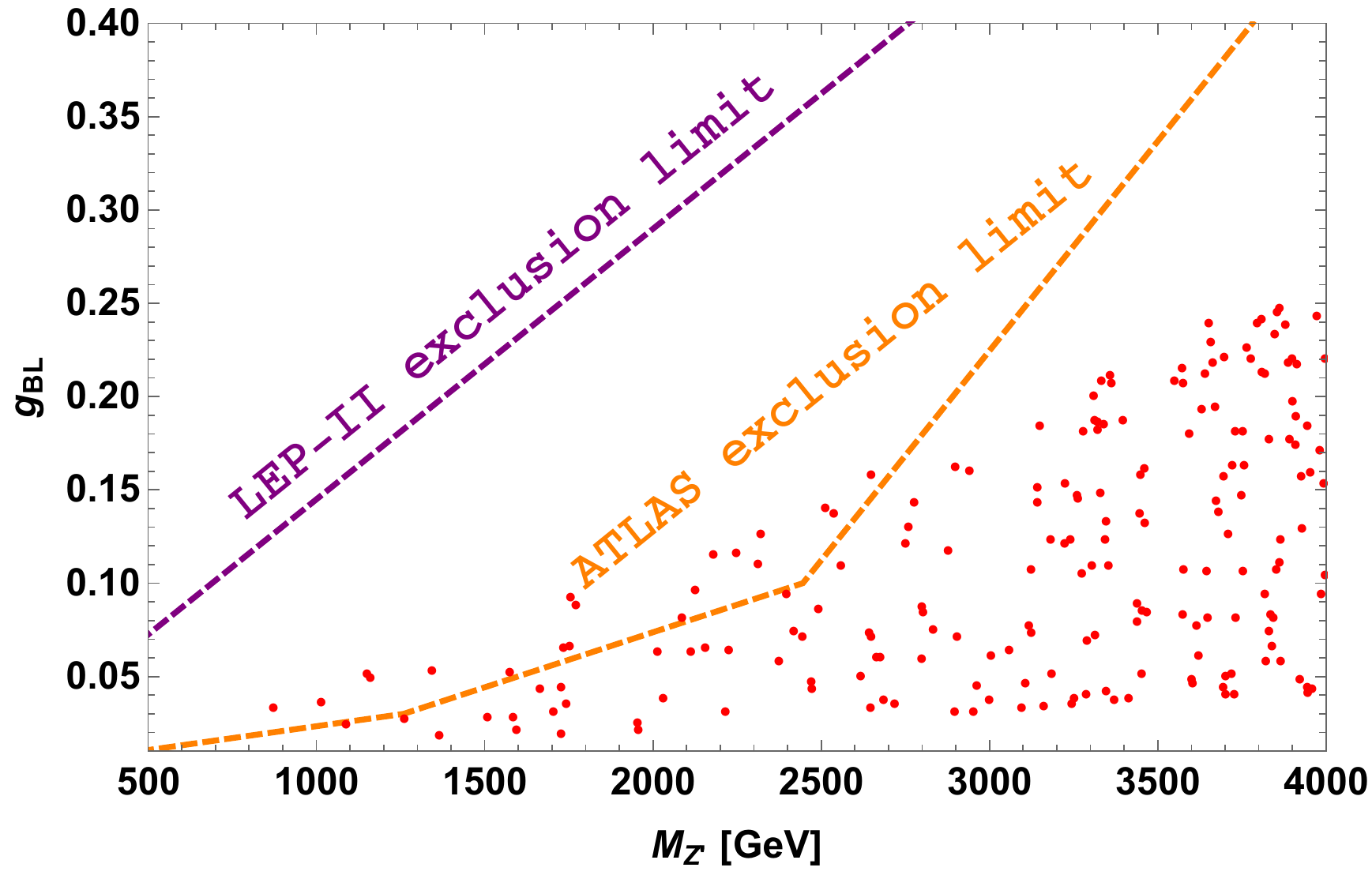}
\caption{Dilepton constraints from ATLAS on the current model are shown here. The black dashed line in the left panel represents the exclusion limit from ATLAS \citep{TheATLAScollaboration:2015jgi}, with the colored lines being the dilepton signal cross sections for various values of $g_{\rm BL}$ as a function of $M_{Z^{\prime}}$. The right panel shows exluclusion limits from LEP-II and ATLAS exclusion limits  in the plane of $M_{Z^{\prime}}- g_{\rm BL}$. The red  points are consistent with the $3\sigma$ range of the relic density limit of PLANCK and the direct detection limits from XENON1T.}
\label{ATLAS}
\end{center}
\end{figure}

The $B-L$ models are further constrained by the collider limits. The extensive analysis of ATLAS and CMS collaborations in search of new heavy resonances in dilepton and dijet signals have derived a lower bound on the mass of $Z^{\prime}$. Recent works \cite{Klasen:2016qux, Okada:2016gsh} have discussed the sensitivity of the bounds from ATLAS \cite{TheATLAScollaboration:2015jgi} on the parameters $M_{Z^{\prime}}$ and $g_{\rm{BL}}$ in $B-L$ models. 
In the present work, we use CalcHEP \cite{Belyaev:2012qa,Kong:2012vg} to compute the production cross section of $Z^{\prime}$. In  the left panel of Fig.
\ref{ATLAS}, we show the $Z^{\prime}$ production cross section times the branching ratio of dilepton ($ee,\mu\mu$) signal  as a function of $M_{Z^{\prime}}$. The black dashed line denotes the dilepton bound from ATLAS  \cite{TheATLAScollaboration:2015jgi}. It is clear that the region below $M_{Z^{\prime}}\simeq3.7$ TeV is excluded for $g_{\rm BL}=0.4$. For   
$g_{\rm BL}=0.1$, $M_{Z_{^\prime}}<2.3$ TeV is ruled out. We have $M_{Z^{\prime}}\gtrsim 1.2$ TeV for $g_{\rm BL} = 0.03$ and the mass region of $M_{Z^{\prime}}\gtrsim 0.5$ TeV is allowed for $g_{\rm BL}=0.01$. The plot in the right panel of Fig. \ref{ATLAS} shows the parameter space that satisfies $3\sigma$ range in the PLANCK relic density limit and the XENON1T constraint. The region to the right of both the dashed curves is consistent with LEP-II \cite{Schael:2013ita} and ATLAS  \cite{TheATLAScollaboration:2015jgi} dilepton limit. We see that the ATLAS gives more stringent limit in the mass region $M_{Z^{\prime}}\lesssim 2.7$ TeV.
\subsection{Landau Pole}
In the present work, we have extra neutral fermions with $B-L$  charges $-4,-4,+5$ required for consistent gauge anomaly free theory. Also there are scalars $\phi_1,\phi_8$ with $B-L$ charges $-1,+8$ and $\phi_{\rm DM}$ with $n_{\rm DM} = 4$. The presence of these additional field content with exotic $B-L$ charges can severely modify the running of the corresponding
gauge coupling $g_{\rm BL}$ and can even lead to a Landau pole $\Lambda_{\rm LP}$. Defining the $U(1)_{B-L}$ fine-structure coupling $\alpha_{\rm BL} = \frac{g^2_{\rm BL}}{4\pi}$, one finds the standard analytic one-loop solution for the renormalization-group running from a scale $\lambda$ to $\Lambda > \lambda$:
\begin{equation}
\frac{1}{\alpha_{\rm BL}(\Lambda)} = \frac{1}{\alpha_{\rm BL}(\lambda)} - \frac{b_{\rm BL}}{2\pi}{\rm{log}}\left(\frac{\Lambda}{\lambda}\right).
\end{equation}
The location of Landau pole $\Lambda_{\rm LP}$ can obtained by $\alpha^{-1}_{\rm BL}(\Lambda_{\rm LP}) = 0$, given by the scale
\begin{equation}
\Lambda_{\rm LP} \simeq \lambda ~{\rm{exp}}\left[\frac{2\pi}{b_{\rm BL}} \alpha^{-1}_{\rm BL}(\lambda) \right],
\end{equation}
for $b_{\rm BL} > 0$. We consider the $B-L$ breaking scale ($\lambda$) and all new particle masses at $10$ TeV. The one-loop beta coefficient, $b_{\rm BL}$ can be computed using the standard formula
\begin{equation}
b_{\rm BL} = -\frac{11}{3}C_2(G) + C_{\rm{norm}}^2\left[\frac{2}{3}\sum_{R_f} \left(\frac{n^f_{\rm BL}}{2}\right)^2 \prod_{j\neq \rm{BL}} d_j(R_f) + \frac{1}{3}\sum_{R_s} \left(\frac{n^s_{\rm BL}}{2}\right)^2 \prod_{j\neq \rm{BL}} d_j(R_s)\right].
\end{equation}
Here $n_{\rm BL}^{f(s)}$ stands for the $B-L$ charge of the  fermion (scalar). $C_2(G)$ denotes the quadratic Casimir operator for gauge bosons which takes the value $N$ for $SU(N)$ and 0 if $U(1)$, $d_j(R_{f,s})$ indicate the dimension of representation $R_{f,s}$ under all the $SU(N)$ groups. Since the model can't be embedded in $SO(10)$ framework, the normalization factor $C_{\rm norm}$ is a free parameter. For instance, choosing $g_{\rm BL} = 0.4$, the Landau pole is at 
\begin{equation}
	\Lambda_{\rm LP} = 
	\begin{cases} 
		5.72 \times 10^{11}~{\rm GeV} & \text{if } C_{\rm norm} = \sqrt{\frac{3}{2}}, \\
    2.48 \times 10^{23}~{\rm GeV} & \text{if } C_{\rm norm} = \sqrt{\frac{3}{5}}, \\
    1.07 \times 10^{35}~{\rm GeV} & \text{if } C_{\rm norm} = \sqrt{\frac{3}{8}}.
	\end{cases}
\end{equation}
Thus, choosing suitable normalization factor, one can avoid the Landau pole below Planck scale.

\section{Scalar portal phenomenology}
\subsection{Relic density}
\begin{figure}[htb]
\begin{center}
\includegraphics[width=0.48\linewidth]{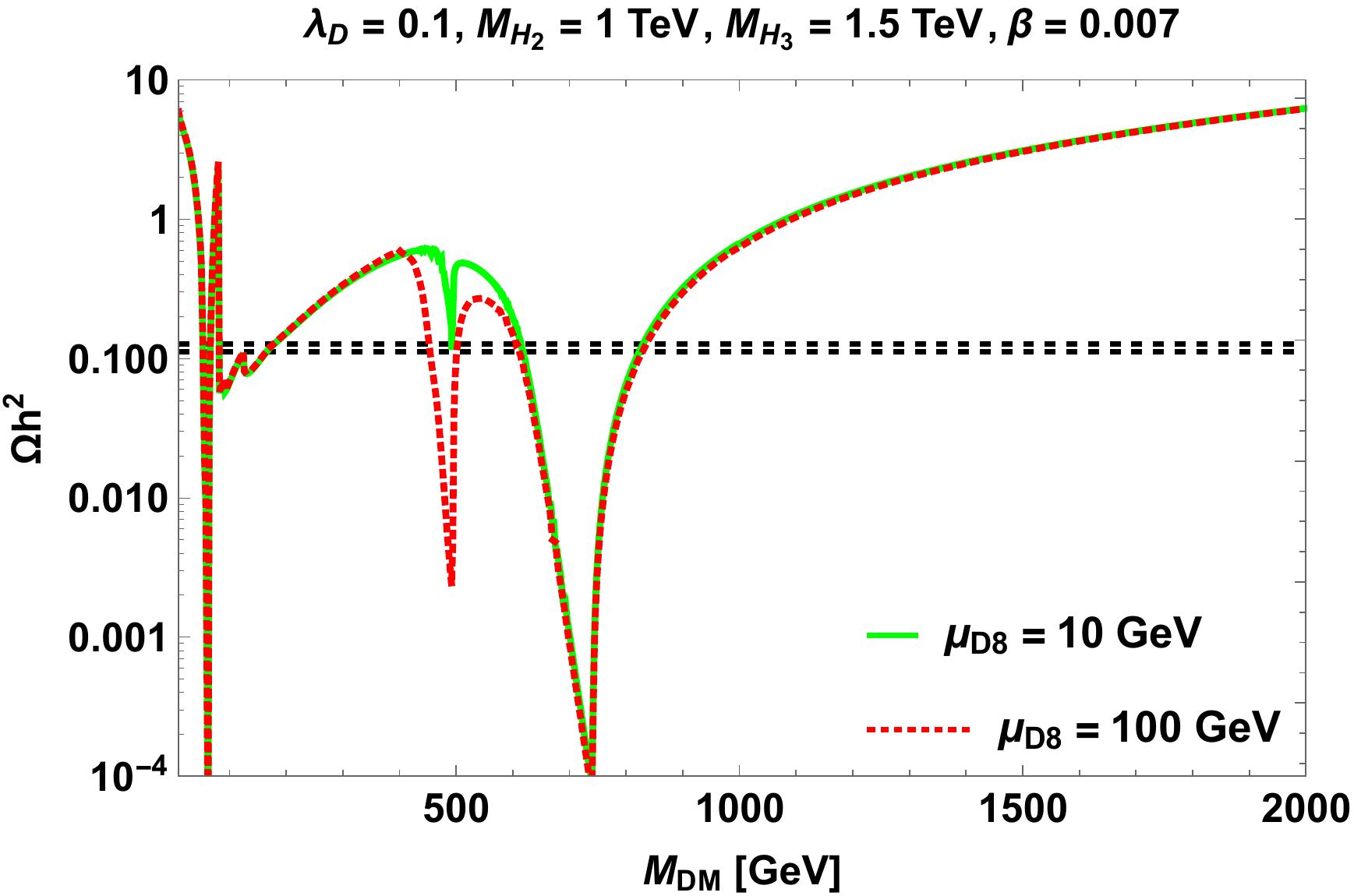}
\caption{Variation of relic abundance $\Omega \text{h}^2$ with the mass of DM for various values of $\mu_{\rm D8}$ parameter. Here the horizontal dashed lines denote the $3 \sigma$ range in current relic density \citep{Ade:2015xua}.}
\label{homega}
\end{center}
\end{figure}

With $n_{\rm DM} = 4$, one can write a non-trivial term to the scalar potential as 
\begin{equation}
\frac{\mu_{\rm D8}}{2} \left[(\phi_{\rm DM})^2 \phi_8^\dagger + (\phi_{\rm DM}^\dagger)^2 \phi_8\right].
\label{muD8}
\end{equation}
The masses of real and imaginary components of $\phi_{\rm DM}$ are given by
\begin{eqnarray}
M_{S}^2 = \mu_{\rm DM}^2 + \frac{ \lambda_{\rm HD}}{2} v^2 +   \frac{ \lambda_{\rm D1}}{2} v^2_1 + \frac{ \lambda_{\rm D8}}{2} v^2_8 + \frac{\mu_{\rm D8} v_8}{\sqrt{2}}, \nn\\
M_{A}^2 = \mu_{\rm DM}^2 + \frac{ \lambda_{\rm HD}}{2} v^2 +   \frac{ \lambda_{\rm D1}}{2} v^2_1 + \frac{ \lambda_{\rm D8}}{2} v^2_8 - \frac{\mu_{\rm D8} v_8}{\sqrt{2}}.
\label{massplit}
\end{eqnarray}
\begin{figure}[t!]
\begin{center}
\includegraphics[width=0.3\linewidth]{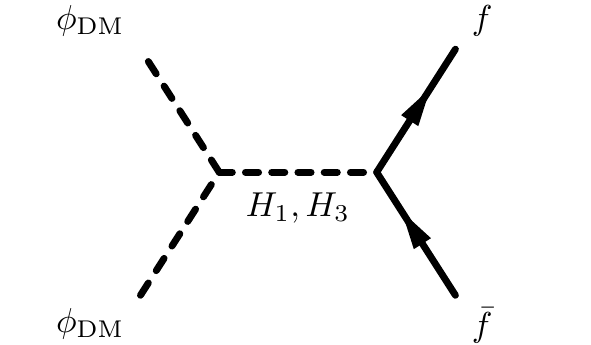}
\includegraphics[width=0.3\linewidth]{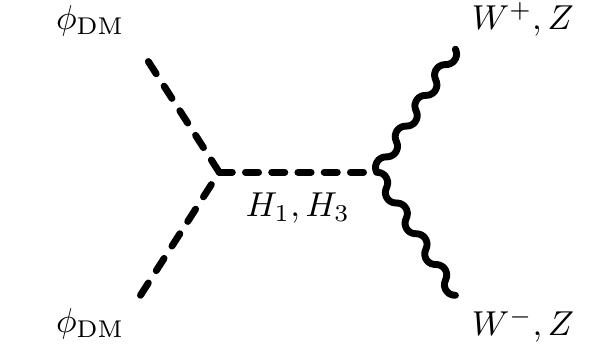}
\includegraphics[width=0.3\linewidth]{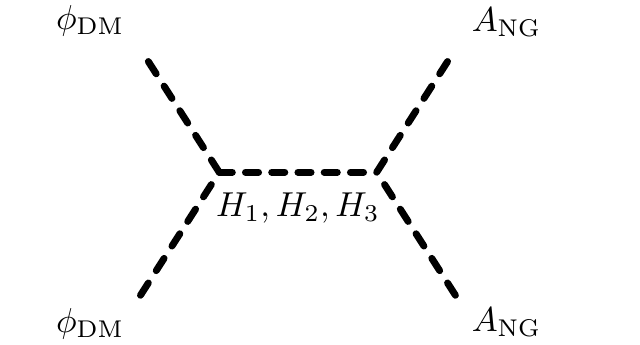}
\includegraphics[width=0.3\linewidth]{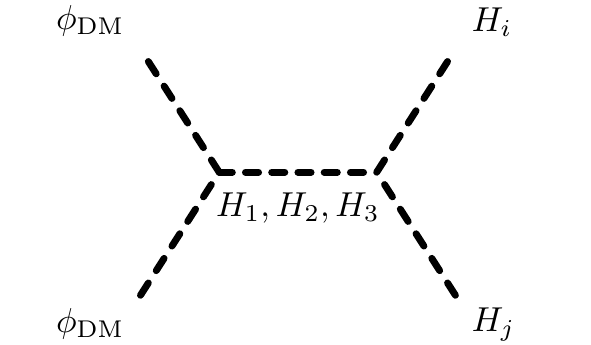}
\includegraphics[width=0.3\linewidth]{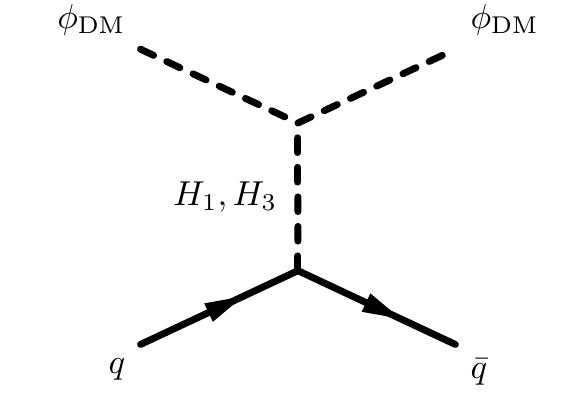}
\caption{Feynman diagrams contributing to relic density in the scalar portal case except the figure of t-channel process is relevant in direct detection studies.}
\label{hfeyn}
\end{center}
\end{figure}
For simplicity, we consider $\lambda_{\rm HD} = \lambda_{\rm H1} = \lambda_{\rm H8} = \lambda_{\rm D}$. The expressions for annihilation cross section of various channels that contribute to relic density shown in Fig. \ref{hfeyn} are
\begin{align}\label{wwann}
	\hat{\sigma}^S_{ff} =& \frac{1}{8\pi v^2s}|F_1|^2
\sum_f M^2_f c_f \frac{(s-4 M^2_{f})^\frac{3}{2}}{(s - 4M^2_{\rm DM})^{\frac{1}{2}}}\;,\\
\hat{\sigma}^S_{WW} =& \frac{s}{16\pi v^2}|F_1|^2
\left(1- \frac{4 M^2_W}{s} + \frac{12 M^4_W}{s^2}\right) \frac{(s-4
	M^2_{W})^\frac{1}{2}}{(s - 4M^2_{\rm DM})^{\frac{1}{2}}}\;,\\
\hat{\sigma}^S_{ZZ} =&\frac{s}{32\pi v^2}|F_1|^2 \left(1- \frac{4 M^2_Z}{s} + \frac{12
M^4_Z}{s^2}\right) \frac{(s-4 M^2_{Z})^\frac{1}{2}}{(s - 4M^2_{\rm DM})^{\frac{1}{2}}}\;,\\
	\hat{\sigma}^S_{\rm{NG}} =&\frac{1}{8\pi} \left(\frac{1}{v_1v_8(v_1^2+64v_8^2)}\right)^2|F_2|^2 
 \frac{s^\frac{3}{2}}{(s - 4M^2_{\rm DM})^{\frac{1}{2}}} \;,\label{zzann}
\end{align}
where
\begin{align*}
	F_1=&\frac{\lambda_{\rm DH1}}{\left[(s-M^2_{H_1}) + i M_{H_1} \Gamma_{H_1}\right]}
- \frac{\sqrt{2}\beta \lambda_{\rm DH3}}{\left[(s-M^2_{H_3}) + i M_{H_3} \Gamma_{H_3}\right]}\;,\\
F_2=&-\frac{\lambda_{\rm DH1}\beta(v_1^3+64v_8^3)}{\left[(s-M^2_{H_1}) + i M_{H_1} \Gamma_{H_1}\right]} +
	\frac{\lambda_{\rm DH2}(v_1^3-64v_8^3)}{\sqrt{2}\left[(s-M^2_{H_2}) + i M_{H_2} \Gamma_{H_2}\right]} - 	\frac{\lambda_{\rm DH3}(v_1^3+64v_8^3)}{\sqrt{2}\left[(s-M^2_{H_3}) + i M_{H_3} \Gamma_{H_3}\right]}\;,
\end{align*}
with $c_f$ and $M_f$ denoting the color charge and mass of the the SM fermion $f$ respectively. Finally, for the Higgs sector annihilation channels  we have
\begin{align*}	
	\hat{\sigma}^S_{H_i H_j} =& \frac{1}{16\pi s n!}|F_{ij}|^2 
 \frac{\left [(s- (M_{H_i}+ M_{H_j})^2)(s- (M_{H_i}- M_{H_j})^2) \right ]^\frac{1}{2}}{\left [s(s - 4M^2_{\rm DM})\right ]^{\frac{1}{2}}}\;,
\label{hann}
\end{align*}
where
\begin{align*}
F_{ij}=&(1+2\beta^2)\lambda_{\rm D}\delta_{ij} + \frac{\lambda_{\rm DH1}\lambda_{1ij}}{\left[(s-M^2_{H_1}) + i M_{H_1} \Gamma_{H_1}\right]} +
	\frac{\lambda_{\rm DH2}\lambda_{2ij}}{\left[(s-M^2_{H_2}) + i M_{H_2} \Gamma_{H_2}\right]} \\
	+&	\frac{\lambda_{\rm DH3}\lambda_{3ij}}{\left[(s-M^2_{H_3}) + i M_{H_3} \Gamma_{H_3}\right]}\;.
\end{align*}
\begin{table}
\begin{center}
\begin{tabular}{|c|c|c|c|c|c|c|c|c|c|}
	\hline
	Coupling & Expression	[GeV] \\
	\hline
	$\lambda_{\rm DH1}$ &$v\lambda_{\rm D} - \frac{1}{4}\beta (8v_1\lambda_{\rm D}-\sqrt{2}\mu_{\rm D8})$   \\
	$\lambda_{\rm DH2}$ &$-\frac{\mu_{\rm D8}}{4}$   \\
    $\lambda_{\rm DH3}$ &$-\sqrt{2}v_1\lambda_{\rm D} - \sqrt{2}v\beta\lambda_{\rm D} + \frac{\mu_{\rm D8}}{4}$   \\	
	\hline
\end{tabular}
\caption{Dark matter couplings to scalars.}
\label{DH}
\end{center}
\end{table}
In all the above expressions $i,j = 1,2,3$, and $\lambda_{\rm DHi}$ denote the coupling of the terms $A^2 H_i$. We show in Fig. \ref{homega}, the scalar portal relic abundance as a function of DM mass. The PLANCK limit on relic density is met near the resonance of three scalar propagators. The channels with $H_1H_1$ and $A_{\rm NG}A_{\rm NG}$ in final state can only give resonance near $M_{\rm DM} \simeq \frac{M_{H_2}}{2}$. However, the coupling $\lambda_{211}$ vanishes. Hence, the channel with NG pair plays a crucial role in giving the resonance in $H_2$ propagator for non-zero $\lambda_{\rm DH2}~ ( = \mu_{\rm D8})$ given in Table. \ref{DH}. One can also notice that the coupling $\mu_{\rm D8}$ induces mass splitting in the scalar components given in Eqn. \ref{massplit}, which is essential to generate light neutrino mass at one loop level to be discussed in the upcoming section.

\subsection{Direct searches}
\begin{figure}[htb]
\begin{center}
\includegraphics[width=0.48\linewidth]{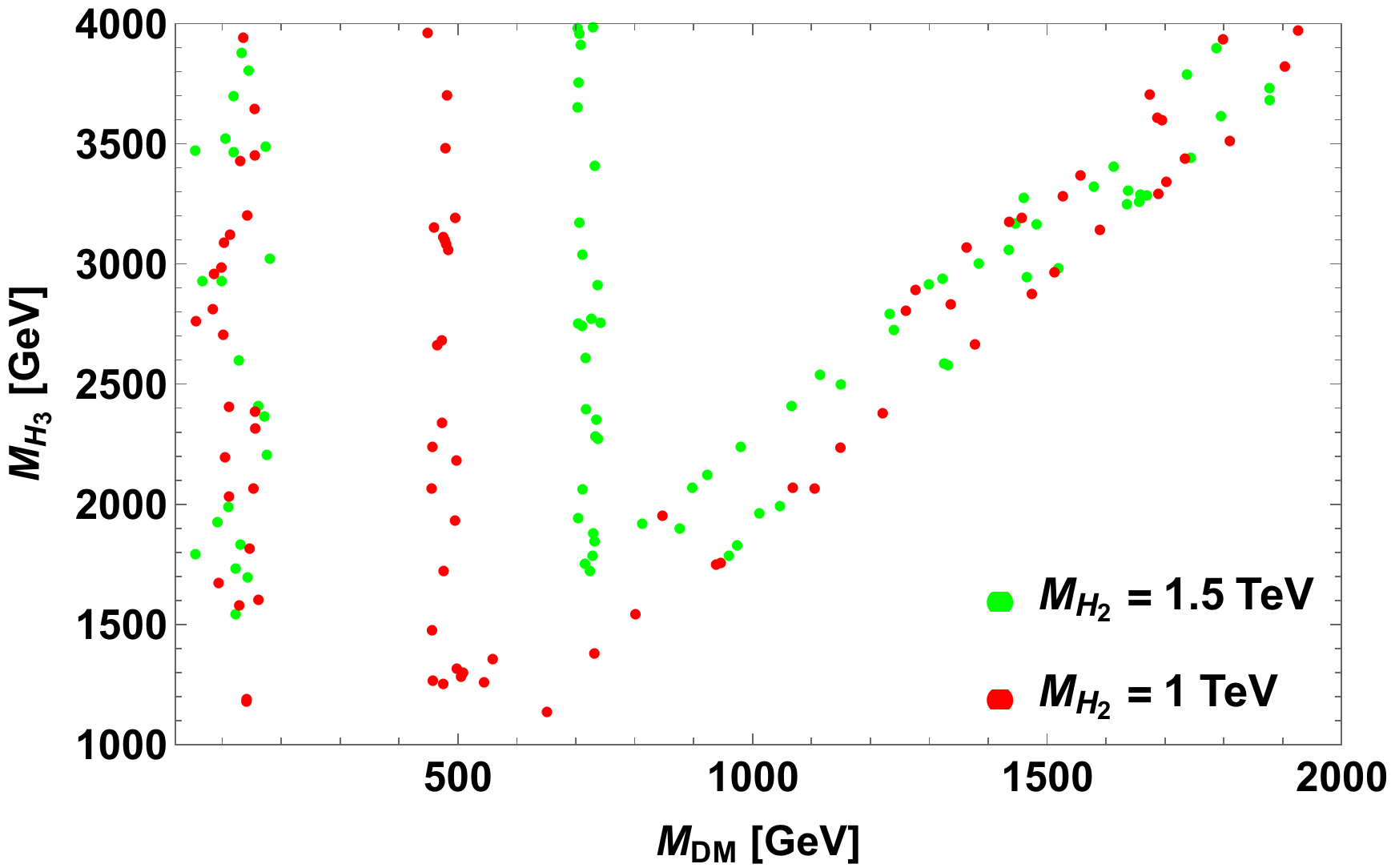}
\vspace{0.2 cm}
\includegraphics[width=0.48\linewidth]{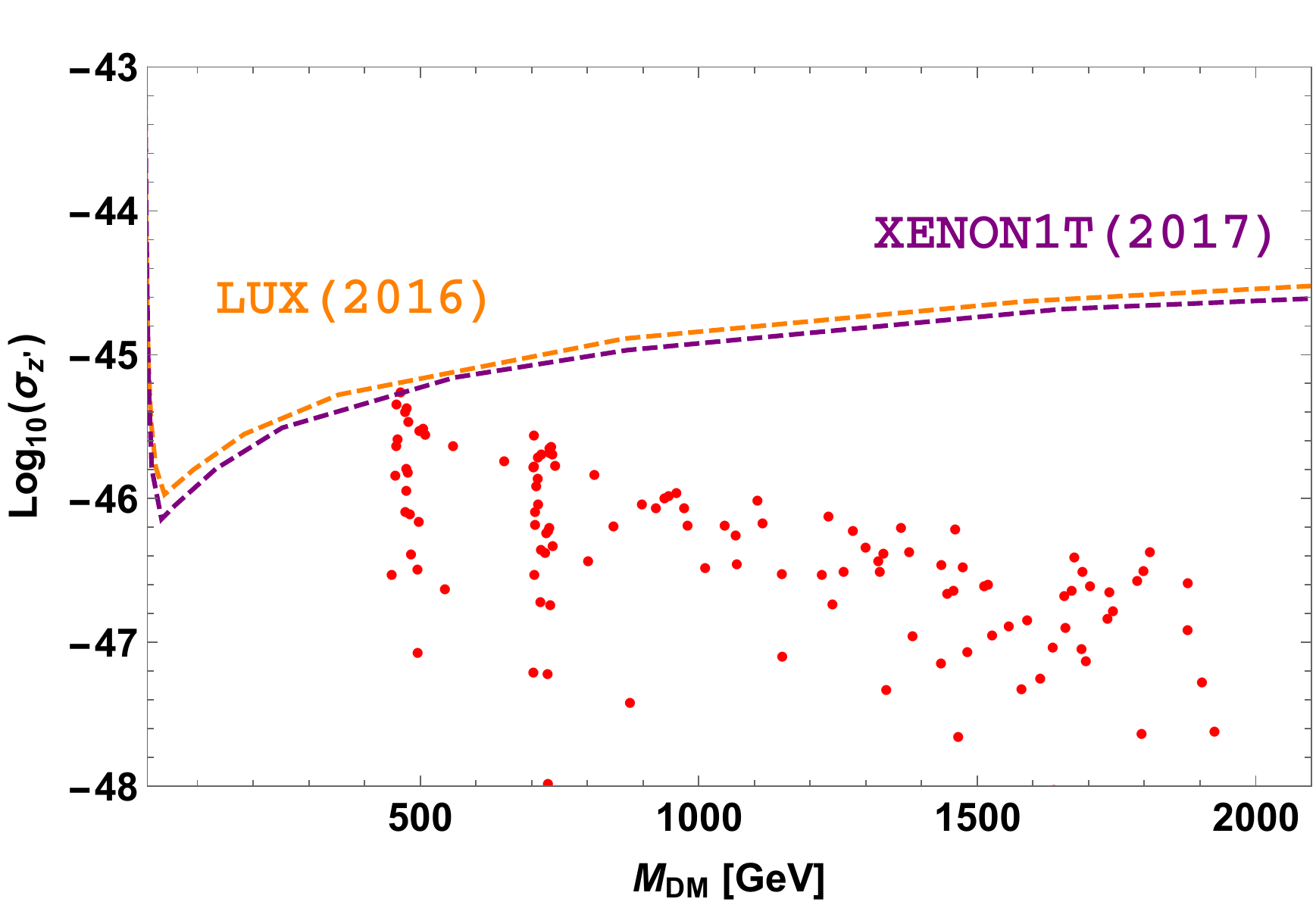}
\caption{Region of $M_{\rm DM}-M_{H_3}$ that meets the $3\sigma$ range on current relic density limit of PLANCK in left panel. Right panel denotes the parameter space taken from left panel that satisfies the XENON1T limit as well. Dashed lines denote the upper limit on WIMP-nucleon cross section by LUX \cite{Akerib:2016vxi} and XENON1T  \cite{Aprile:2017iyp}.}
\label{hdirect}
\end{center}
\end{figure}

In the scalar portal scenario, the DM-WIMP nucleon cross section is given by
\begin{equation}
\sigma_{S}=\frac{\mu^2 M_{n}^2}{4\pi v^2 M_{\rm DM}^2}\left[\frac{\lambda_{\rm DH1}}{M_{H_1}^2} - \frac{\sqrt{2}\lambda_{\rm DH3}\beta}{M_{H_3}^2} \right]^2 f_p^2 ,
\end{equation}
where $M_n$ is the nucleon mass, $\mu$  denotes the reduced mass of WIMP-nucleon system and $f_p$ is given by
\begin{equation}
f_p=\frac{2}{9}+\frac{7}{9}\sum_{q=u,d,s}f^{p}_{Tq}.
\end{equation}
Typical values for proton are $f_{Tu}^{p} = 0.020\pm 0.004$,  $f_{Td}^{p} = 0.026\pm0.005$
and $f_{Ts}^{p} = 0.118\pm0.062$ \cite{Ellis:2000ds}. Varying the model parameters given in Table. \ref{paraH}, Fig. \ref{hdirect} left panel denotes the parameter space in $M_{H_3}-M_{\rm DM}$ plane satisfying $3\sigma$ range on current relic density limit by PLANCK and the right panel denotes the  allowed parameter space (corresponding to the allowed parameters of the left panel), consistent with XENON1T limit. We see that the data points near the resonance of SM Higgs $H_1$ doesn't satisfy the XENON1T limit on WIMP-nucleon cross section.
\begin{table}[htb]
\begin{center}
\begin{tabular}{|c|c|}
	\hline
	Parameters & Range  \\\hline
	$\mu_{\rm D8}$ [GeV]& $10-100$ \\
	$\lambda_{\rm D}$ & $0.001 - 0.1$ \\
	$v_{1,8}$ [GeV] & $2000$\\
	$M_{H_{2}}$ [GeV]  & $1000,2000$\\
	$M_{H_{3}}$ [GeV]  & $M_{H_2}-4000$\\
	$M_{\rm DM}$ [GeV]  & $20-2000$\\
	$\beta$ & $0.001-0.016$ \\
	\hline
	\hline
\end{tabular}
\caption{Parameters and their ranges for scalar portal analysis.}
\label{paraH}
\end{center}
\end{table}
\section{Light neutrino mass}
\begin{figure}[htb]
\begin{center}
\includegraphics[width=7.5 cm,height= 5.0 cm, clip]{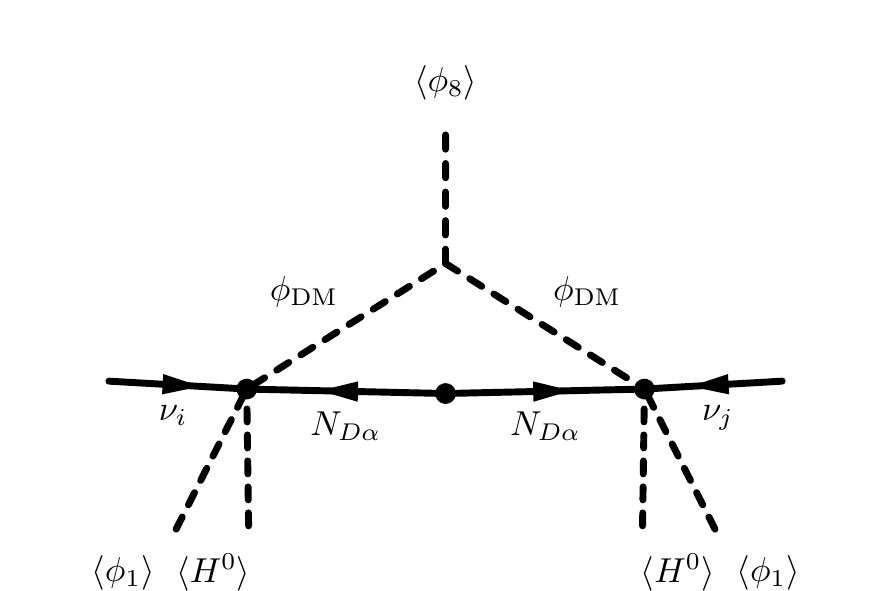}
\caption{Radiative generation of neutrino mass}
\label{rad_nu}
\end{center}
\end{figure}

The light neutrino mass in this model can be achieved by radiative mechanism.   The model structure permits us to write a dim-6 Yukawa interaction term of the form\footnote{The interaction term in Eqn. \ref{radnumass} can induce the decay $\nu_i \to \nu_j + A_{\rm NG}$. However the decay rate of this channel can be greater than the age of the universe \cite{Schechter:1981cv}. The effect of neutrino decay in neutrino oscillations has been investigated in literature \cite{Lindner:2001fx}. However, this study is beyond the scope of our work.}
\begin{equation}
\frac{1}{\Lambda^2}\sum_{\alpha=1,2} Y_{i\alpha} \overline{(\ell_L)}_i \tilde{H} N_{\alpha R}\phi_{\rm DM} \phi_1.\label{radnumass}
\end{equation}
Now, it is possible to generate the light neutrino mass at one loop level as shown in Fig. \ref{rad_nu}.
If we assume the masses of real and imaginary components of $\phi_{\rm DM}$ satisfy the relation $(M^2_S + M^2_A)/{2} \gg M^2_S -M^2_A = \sqrt{2} \mu_{\rm D8} v_8$, the expression for the radiatively generated neutrino mass is \cite{Ma:2006km}
\begin{equation}
({\cal M}_\nu)_{ij} = {\sqrt{2}\mu_{\rm D8}v_8 v^2 v^2_1 \over16 \pi^{2} \Lambda^4} 
\sum_{\alpha=1}^3 {Y_{i \alpha} Y_{j\alpha} M_{D\alpha} \over m_0^{2} - M_{D\alpha}^{2}} \left[ 
1 - {M_{D\alpha}^{2} \over m_0^{2}-M_{D\alpha}^{2}} \ln {m_0^{2} \over M_{D\alpha}^{2}}  \right],
\end{equation}
where we denote  $m_0^2 = (M^2_S + M^2_A)/{2}$. If $ M_{D\alpha}^2 \gg m^2_0$, then
\begin{equation}
({\cal M}_\nu)_{ij} = {\sqrt{2} \mu_{\rm D8}v_8 v^2 v^2_1 \over 16 \pi^2 \Lambda^4} 
\sum_{\alpha=1}^3 {Y_{i \alpha} Y_{j\alpha} \over M_{D\alpha}}\left[ \ln\frac{M^2_{D\alpha}}{m^2_0}-1\right].
\end{equation}
Here
$M_{D\alpha} = (U^T M U)_{\alpha}$ and $N_{D\alpha} = U^{\dagger}_{\alpha \beta} N_{\beta}$, with $M$ being the Majorana mass matrix. 
Assuming the lightest exotic fermion gives dominant contribution to the light neutrino mass matrix and considering $(v_1,v_8,M_{D\alpha_1}) \sim (2,2,3)$ TeV ($M_{D\alpha_1}$ being the mass of the lightest exotic fermion mass eigenstate), we show sample benchmark values  in Table. \ref{numass_bench} that satisfy PLANCK, XENON1T limit and $\nu$-mass simultaneously. We conclude that this model is quite advantageous to explain the light neutrino mass even without the small Yukawa couplings.
\begin{table}
\begin{center}
\begin{tabular}{|c|c|c|c|c|c|c|c|c|c|}
	\hline
	Sl. No. & $\mu_{\rm D8}$~&~ $\sqrt{|Y_{i \alpha_1 } Y_{j \alpha_1}|}$~ & ~$M_{\rm DM}$ [GeV]~&~$\Lambda$ [TeV]~& $m_{\nu}$ [GeV]&$\Omega h^2$~&~$\rm{Log}_{10}\sigma_S ~ [\rm{cm^2}]$\\
	\hline
	$1.$ & $96.8$ & $0.026$ & $448$ &$50$& $4.01\times 10^{-11}$ &$0.1154$&$-46.517$\\	
	$2.$ & $21.3$& $0.056$ & $483$ & $50$ &$4.09\times 10^{-11}$&$0.123$ &$-46.376$\\
$3.$ & $21.3$& $0.22$ & $483$ & $100$ &$3.95\times 10^{-11}$ &$0.123$ &$-46.376$\\
	\hline
\end{tabular}
\caption{Sample benchmark for radiative $\nu$-mass.}
\label{numass_bench}
\end{center}
\end{table}
\section{Semi-annihilations for Scalar Dark matter}
\begin{figure}[h!]
\centering
\includegraphics[width=0.3\linewidth]{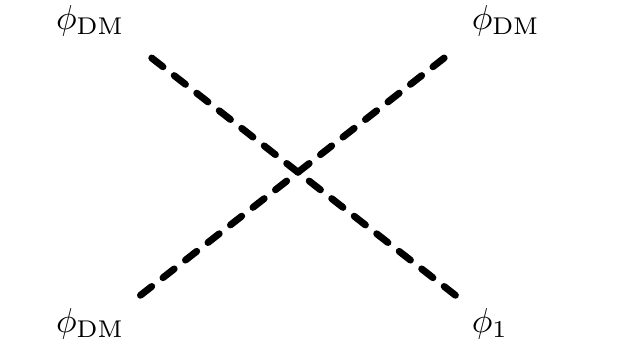}
\caption{Feynman diagram for the semi-annihilation vertex.}
\label{feyn:semi}
\end{figure}

Fractional $B-L$ charge to the inert scalar can induce semi-annihilations which can show up in dark matter relic abundance (see Refs.\cite{DEramo:2010keq,Rodejohann:2015lca}). For instance when $n_{\rm DM} = 1/3$, 
there is a quartic term in the Lagrangian of the form
\begin{equation}
\mathcal{L}_{1/3} = \frac{\lambda^\prime_{\rm DM}}{3} \phi_{\rm DM}^3 \phi_1 + \text{h.c.}
\end{equation} 
With the Feynman diagram shown in Fig. \ref{feyn:semi}, the cross section of all possible semi-annihilation channels are
\begin{eqnarray}
\hat{\sigma}^{1/3}_{H_1} &=&  \frac{ {\lambda^{\prime}_{\rm DM}}^2\beta^2}  {64 \pi s } \frac{\left[(s- (M_{\rm DM} + M_{H_1})^2)(s- (M_{\rm DM} - M_{H_1})^2)\right]^{\frac{1}{2}}}{[s(s-4 M^2_{\rm DM})]^{\frac{1}{2}}}\,,\nn\\
\hat{\sigma}^{1/3}_{H_2} &=&  \frac{ {\lambda^{\prime}_{\rm DM}}^2}  {128 \pi s } \frac{\left[(s- (M_{\rm DM} + M_{H_2})^2)(s- (M_{\rm DM} - M_{H_2})^2)\right]^{\frac{1}{2}}}{[s(s-4 M^2_{\rm DM})]^{\frac{1}{2}}}\,,\nn\\
\hat{\sigma}^{1/3}_{H_3} &=&  \frac{ {\lambda^{\prime}_{\rm DM}}^2}  {128 \pi s } \frac{\left[(s- (M_{\rm DM} + M_{H_3})^2)(s- (M_{\rm DM} - M_{H_3})^2)\right]^{\frac{1}{2}}}{[s(s-4 M^2_{\rm DM})]^{\frac{1}{2}}}\,.
\end{eqnarray}
\begin{figure}[htb]
\begin{center}
\includegraphics[width=7.5 cm,height= 5.0 cm, clip]{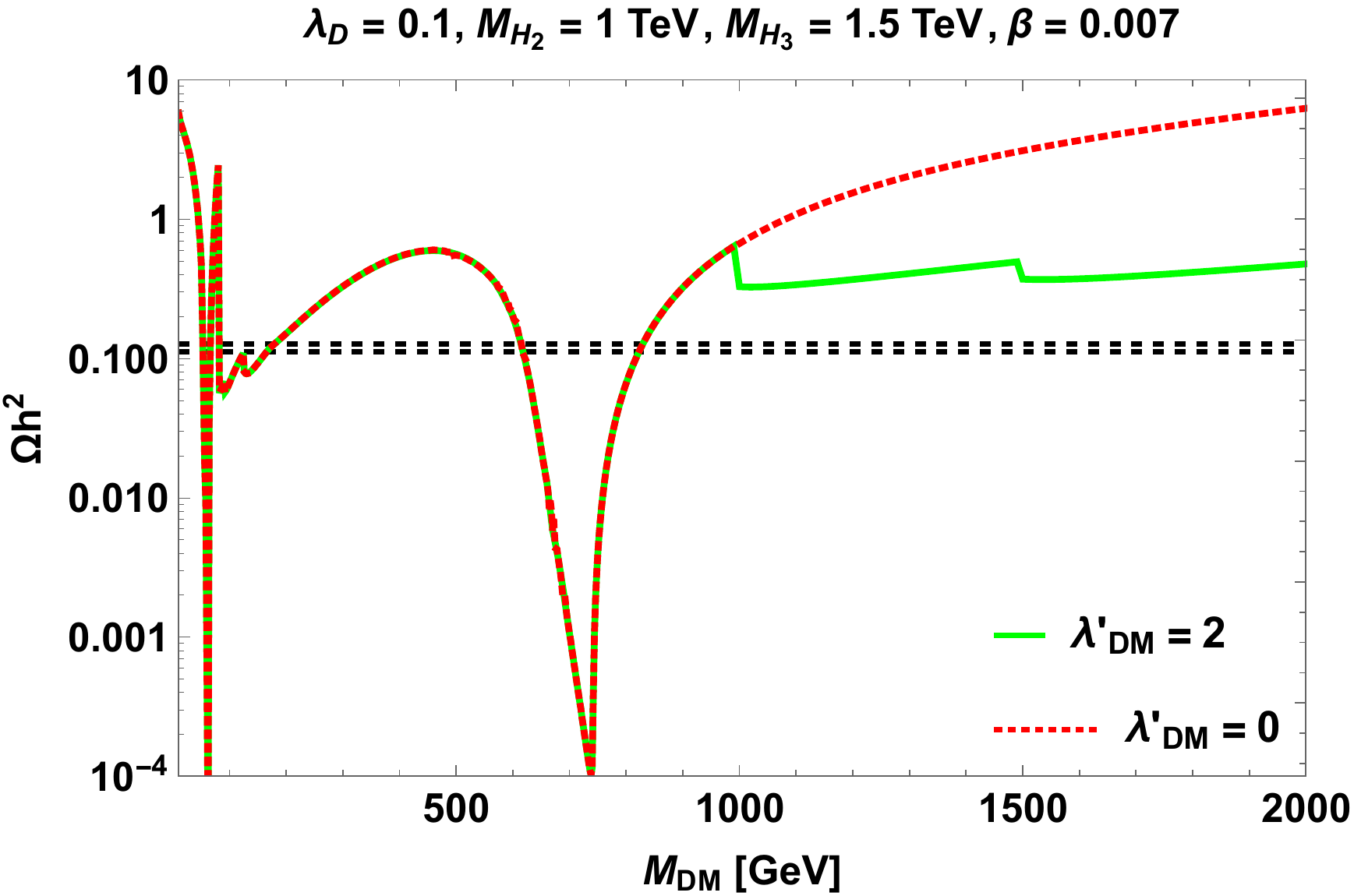}
\caption{Relic abundance $\Omega \text{h}^2$ with the mass of DM plotted for two values of $\lambda^{\prime}_{\rm DM}$ with the choice of $n_{\rm DM} = 1/3$.}
\label{semi_relic}
\end{center}
\end{figure}
We display in Fig. \ref{semi_relic}  the effect of semi-annihilation channel on the relic abundance observable. Resonance near $M_{\rm DM} \simeq \frac{M_{H_2}}{2}$ is not achieved as the term in Eqn. \ref{muD8} doesn't exist for $n_{\rm DM} = 1/3$, in turn making $\lambda_{\rm DH2}$ to be zero. These new channels begin to pop up once mass of DM is above the mass of the physical scalar appearing in the final state. We see that the channel with $H_2$ and $H_3$ as one of the final state particles have a significant effect while the Higgs channel attains a $\beta^2$ suppression. This scenario is very appealing as the dark matter phenomenology is determined by three free parameters i.e, $\lambda^\prime_{\rm DM}$,  $M_{\rm DM}$ and the mass of the physical scalar. Similarly, one can also perform the same analysis for $n_{\rm DM} = 8/3$ as well.
\section{Comment on Indirect signals}
No indirect signals are expected in $Z^{\prime}$-portal scenario as the annihilation rate today is velocity suppressed \cite{Rodejohann:2015lca}. Moving to the scalar-portal, we briefly comment as follows.
\subsection{Gamma ray excess}
Fermi-LAT report of excess in $\gamma$-ray emission, appearing as a peak around $1-3$ GeV energy range  \cite{Calore:2014nla, Daylan:2014rsa} can be well fitted by a DM maximally annihilating to $b\bar{b}$ channel with the mass range $35-165$ GeV and $\langle \sigma v \rangle = (1-3) \times 10^{-26} ~{\rm{cm}}^3 \rm{s}^{-1}$ \citep{Agrawal:2014oha}.
In the current model, the only way that to explain this excess is near the Higgs resonance where the relic density is met near $M_{\rm DM} \simeq \frac{M_{H_1}}{2}$. However, from Fig. \ref{hdirect}, the data points satisfying PLANCK limit (near Higgs resonance) in the left panel are eliminated by the direct detection bounds of XENON1T and LUX, conveyed in the right panel. Hence, the present model does not accommodate the excess $\gamma$-ray emission of Fermi-LAT.
\subsection{Positron excess}
The excess in positron signal reported by PAMELA \cite{Adriani:2013uda} and AMS-02 \cite{Aguilar:2014mma} can be well explained by a DM with mass $350$ and $894$ GeV annihilating to ($e^+e^-$ or $\mu^+\mu^-$) and $\tau^+\tau^-$ respectively. Another possibility is with a two scalar final state channel that subsequently decays to charged lepton pairs for the DM masses $350$ and $590$ GeV \cite{Boudaud:2014dta}. However, at all these mass values of DM, the channel with W-boson pair maximally contributes to the scalar-portal relic density. Therefore, we don't get a proper fit to the observed excess in this model dependent framework.
\section{Conclusion}
In this article, we have presented in detail the scalar dark matter phenomenology in the context of an anomaly free $U(1)_{B-L}$ extension of SM. A possible solution to cancel out the resulting non-trivial triangle anomalies of the gauge extension, three heavy neutral fermions $N_{iR} ~ (i=1, 2, 3)$ with $B-L$ charges $-4, -4$ and $+5$ are added to the existing lepton content of the standard model. Furthermore, the scalar sector is enriched with two scalar singlets $\phi_1$ and $\phi_8$  to spontaneously break the $U(1)_{B-L}$ gauge symmetry and also to provide the Majorana mass terms for the newly added fermions $N_{iR}$. A scalar singlet $\phi_{\rm DM}$ is introduced such that the $U(1)_{B-L}$ symmetry takes the burden to forbid its decay making it a stable dark matter candidate. Three physical scalars and a heavy gauge boson $Z^{\prime},$ a resultant of having $U(1)_{B-L}$ as local gauge symmetry act as mediators between the visible and dark sector. 

We have studied the scalar spectrum emphasizing the minimization conditions, vacuum stability, perturbative unitarity conditions and their acquired masses after spontaneous symmetry breaking of $SU(2)_L \times U(1)_Y \times U(1)_{B-L}$ gauge symmetry.
Choosing a particular ${B-L}$ charge that can stabilize $\phi_{\rm DM}$, we have investigated thoroughly the relic density and of scalar singlet dark matter in the $Z^{\prime}$ and scalar-portal scenarios. Applying the limits on relic density by PLANCK and the most stringent bounds on WIMP-nucleon spin-independent cross section by LUX and XENON1T, we have obtained the consistent parameter space. In collider studies, we have used ATLAS dilepton limits on the gauge coupling $g_{\rm BL}$ and the mass of the new vector boson $M_{Z^{\prime}}$. We found that there is enough region for the model parameters to meet all the experimental bounds.  

This remarkable gauge extension is economical in particle content and rich in phenomenology. A unique feature of this model is that a massless physical Goldstone boson, which  plays a key role in scalar-portal relic density. We have discussed the mechanism  to obtain the light neutrino mass at one-loop level, with the dark matter singlet running in the loop, and a suitable benchmark, where the dark matter observables and light neutrino mass are simultaneously consistent.
We have included discussions regarding semi-annihilations of dark matter and its imprint on relic density for a choice of fractional $B-L$ charge for the scalar dark matter. We finally commented on indirect signals in the present model. To conclude, the explored model is quite consistent with current bounds of recent and ongoing dark matter experiments and a testable framework built based on the well-tested local gauge principles of Standard Model.\\

{\bf ACKNOWLEDGMENT} 
\vspace*{0.1 true in}

SS would like to thank Dr. Soumya Rao for the help in micrOMEGAs code and Department of Science and Technology (DST) - Inspire Fellowship division, Govt of India for the financial support through ID No. IF130927.  RM would like to thank Science and Engineering Research Board (SERB), Government
of India for financial support through grant No. SB/S2/HEP-017/2013. SP would like to acknowledge the warm hospitality provided by University of Hyderabad, India, between $22^{nd} - 29^{th}$ March,  2017,
during which this work was completed.

\appendix
\section{Massless Goldstone interaction terms}\label{apdx}
\subsection{Exotic fermion sector}
The massless mode, $A_{\rm NG}$ can couple to the heavy fermions by the Majorana mass term in eqn. \ref{dirac} as
\begin{equation}
\frac{iA_{\rm NG}}{\sqrt{v_1^2 + 64v_8^2}}\left[\sum_{\alpha=1,2} \frac{8v_8 h_{\alpha 3} }{\sqrt{2}} \overline{N^c_{\alpha R}} N_{3R}  + \sum_{\alpha, \beta=1,2} \frac{v_1 h_{\alpha \beta}}{\sqrt{2}} \overline{N^c_{\alpha R}}   N_{\beta R}\right],
\end{equation}
where the Majorana mass matrix takes the form
\begin{align}
	M
	=
	\begin{pmatrix}
		h_{11} \langle \phi_8 \rangle	& h_{12} \langle \phi_8 \rangle	& h_{13} \langle \phi_1 \rangle	\\
		h_{12} \langle \phi_8 \rangle	& h_{22} \langle \phi_8 \rangle	& h_{23} \langle \phi_1 \rangle	\\
		h_{13} \langle \phi_1 \rangle	& h_{23} \langle \phi_1 \rangle	& 0				\\
	\end{pmatrix} \;.
	\label{massMatrix1}
\end{align}
Writing it in a simplified form as
\begin{align}
	M
	=
	\begin{pmatrix}
		x~	&~ a~& ~b	\\
		a~	&~ x~	& ~b	\\
		b~	&~ b~	& ~0				\\
	\end{pmatrix}, 
\end{align}
which can be obtained assuming the Yukawa couplings to satisfy the relation $h_{11} \approx h_{22}$ and $y_{13} \approx y_{23}$ along with $v_1 \approx v_8$.  The mass matrix  can be diagonalized by using normalized eigenvector matrix as $M_{D\alpha} = (U^T M U)_\alpha$, and the mass eigenstates are given as $N_{D\alpha} = U^\dagger_{\alpha\beta} N_\beta$.
\subsection{Scalar sector}
The coupling of $A_{\rm NG}$   to the  new scalar fields is given   as 
\begin{equation}
\frac{2} {v_1v_8(v_1^2 + 64v_8^2)} (\partial_\mu A_{\rm NG})^2\left[-H_1\beta\left(v_1^3 + 64 v_8^3 \right) + \frac{H_2}{\sqrt{2}}\left(v_1^3 - 64 v_8^3 \right) - \frac{H_3}{\sqrt{2}}\left(v_1^3 + 64 v_8^3 \right)\right].
\end{equation}

\bibliographystyle{utcaps_mod}
\bibliography{BL.bib}

\end{document}